\documentclass[pra,twocolumn,showpacs,amsmath,nobibnotes,nofootinbib,floatfix,aps,10pt,superscriptaddress]{revtex4-1}
\usepackage{color}
\usepackage{ulem}
\usepackage{graphicx}
\usepackage{epstopdf}
\usepackage{bm}
\usepackage{amssymb}
\usepackage{amsmath}
\usepackage[colorlinks,citecolor=blue]{hyperref}
\usepackage{epsfig}

\def \be {\begin{equation}}
\def \ee {\end{equation}}
\def \bea {\begin{align}}
\def \eea {\end{align}}

\def \BEA {\begin{eqnarray}}
\def \EEA {\end{eqnarray}}

\def \BC {\begin{cases}}
\def \EC {\end{cases}}

\begin{document}

\title{Weak antilocalization in two-dimensional systems with large Rashba splitting}

\author{L.\,E.\,Golub}
\affiliation{Ioffe Institute, St.~Petersburg 194021, Russia}

\author{I.\,V.\,Gornyi}
\affiliation{Ioffe Institute, St.~Petersburg 194021, Russia}
\affiliation{Institut f\"ur Nanotechnologie, Karlsruhe Institute of Technology, 76021 Karlsruhe, Germany}

\author{V.\,Yu.\,Kachorovskii}
\affiliation{Ioffe Institute, St.~Petersburg 194021, Russia}


\begin{abstract}
We develop the theory of quantum transport and  magnetoconductivity 
for two-dimensional electrons with an arbitrary large (even exceeding the Fermi energy), linear-in-momentum Rashba or Dresselhaus spin-orbit splitting. For short-range disorder potential, we derive the analytical expression for the quantum conductivity correction, which accounts for interference processes with an arbitrary number of scattering events and is valid beyond the diffusion approximation.  We demonstrate that the zero-field conductivity correction is given by the sum of the universal logarithmic ``diffusive'' term and a ``ballistic'' term.  The latter is temperature independent and encodes information about spectrum properties. This information can be extracted experimentally  by measuring the  conductivity correction at different temperatures and electron concentrations. We calculate the quantum correction in the whole range of classically weak magnetic fields and find that the magnetoconductivity  is negative both in the diffusive and in the ballistic regimes, for an arbitrary relation between the Fermi energy and the spin-orbit splitting.  We also demonstrate  that the magnetoconductivity changes with the Fermi energy when the Fermi level is above the ``Dirac point'' and 
does not depend on the Fermi energy when it
goes below this point.
\end{abstract}


\maketitle

\section{Introduction}

Weak localization is a coherent phenomenon in the low-temperature  transport  in disordered systems.
Transport in such systems is realized by various trajectories, including a special class of trajectories with closed loops.
The underlying physics of the weak localization is the enhancement of the backscattering amplitude
which results from the constructive interference of the waves propagating along the loops
in the opposite directions (clockwise and counterclockwise).
Since interference increases the backscattering amplitude, quantum conductivity correction is negative and is proportional to the ratio of the de~Broglie wavelength to the mean free path~\cite{AAreview}.   Remarkably, this correction diverges
logarithmically at low temperatures in the two-dimensional (2D) case.
Such a divergence is a precursor of the strong localization and reflects universal symmetry properties of the system.

Dephasing processes suppress interference and, consequently, strongly affect the conductivity correction.
Specifically, the typical size of interfering paths is limited by the time of electron dephasing, $\tau_\phi$.
At low temperatures, the dephasing rate is dominated by inelastic electron-electron collisions. The phase space
for such collisions decreases with lowering temperature.
Therefore, one can probe dephasing processes by  measuring  the temperature dependence  of conductivity in the weak-localization regime~\cite{Bergmann}.
Another possibility to affect the interference-induced quantum correction to the conductivity is application of magnetic field.
The Aharonov-Bohm effect introduces a phase difference for the waves traveling along the closed loop in the opposite directions.
This phase difference equals to a double magnetic flux passing through the loop. The anomalous magnetoconductivity
allows one to extract the dephasing time even more accurately than the temperature measurements since the low-field
magnetoconductivity is not masked by other effects~\cite{AAreview,Gantmakher}.

\begin{figure}[t]
\includegraphics[width=\linewidth]{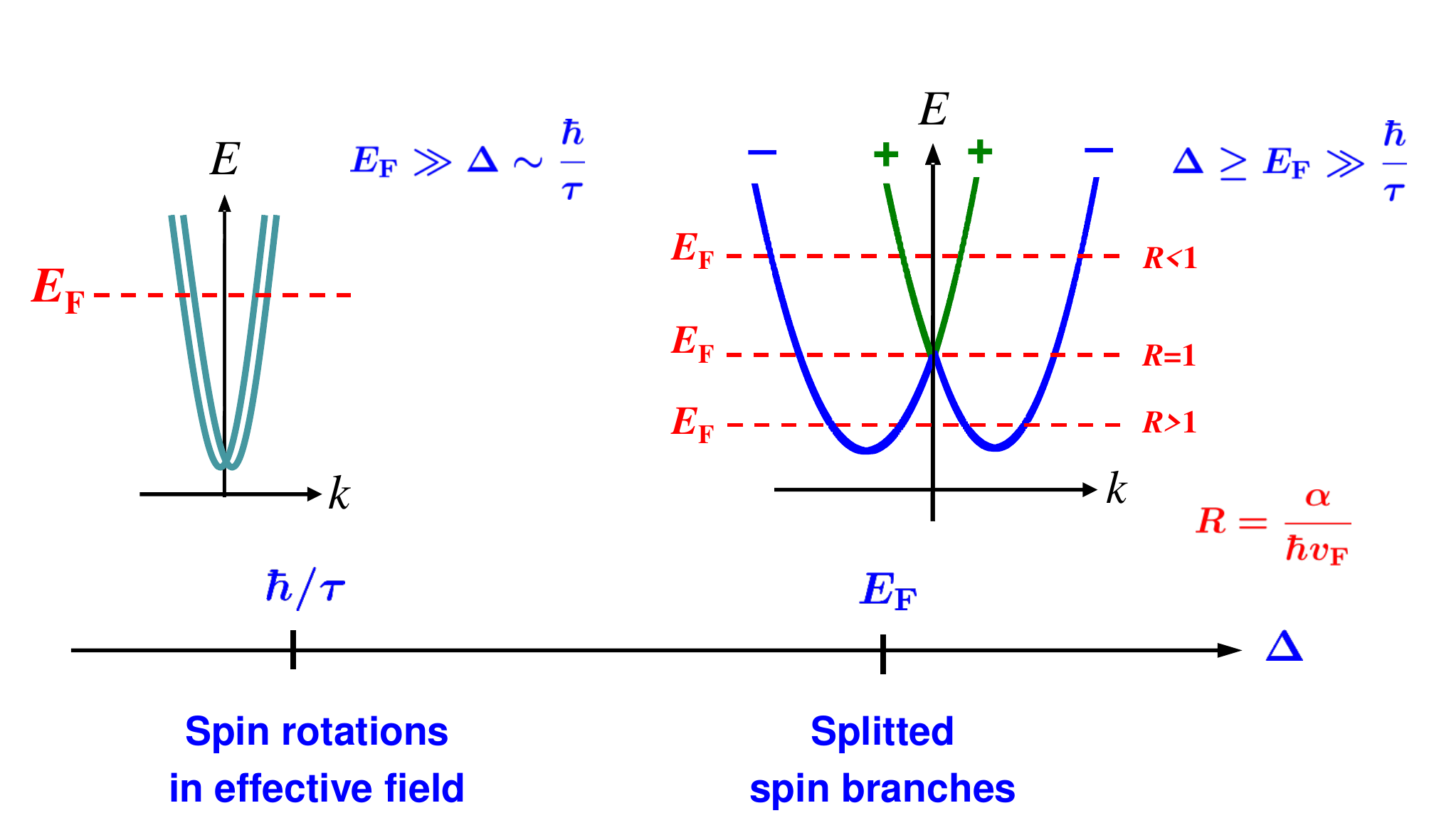}
\caption{ Electron spectrum and regimes of spin dynamics depending on the strength of the Rashba splitting.  Left: Regime of spin rotations in the effective magnetic field realized at moderate spin-orbit splitting $\Delta\sim \hbar/\tau$;  the weak localization theory is developed in Ref.~\cite{GlazovGolub_2008}.
Right: Regime of well separated spin branches at strong spin-orbit splitting ($\Delta\sim E_{\rm F}$) considered in the present work.}
\label{fig:regimes}
\end{figure}

Since weak localization is caused by the interference of paths related to each other by time inversion, it is extremely sensitive to spin properties of interfering particles. In systems with spin-orbit coupling (see Fig.~\ref{fig:regimes}), an additional spin-dependent phase is acquired by electrons passing the loops clock- and anti-clockwise. As a result, the interference depends on the electron spin states before and after passing the loop. Importantly, in the presence of spin-orbit coupling, the interference becomes destructive, resulting in a positive correction to the conductivity. This interference effect is called weak antilocalization. Magnetic field suppresses this correction making the conductivity smaller than in zero field, i.e. the magnetoconductivity is negative~\cite{AAreview}.

Theory of weak localization developed in 1980's for diffusive systems allowed one to explain a number
of experimental data in various metallic and semiconductor structures~\cite{Bergmann}. Spin-orbit interaction has been treated as spin relaxation which adds an additional channel for dephasing of the triplet contributions to the quantum corrections~\cite{AAreview}. However, this approach is insufficient for 2D semiconductor heterostructures with the linear-in-momentum spin-orbit  splitting of the spectrum. 
The relevant theory of weak localization has been developed in the middle of 1990's~\cite{ILP}. It describes very well experimental data~\cite{Knap,tilted}.

With increasing the magnetic field, the magnetic length $l_B$  becomes smaller than the mean free path $l$. 
This regime of weak localization can not be treated within the model of a diffusive electron motion along large scattering paths. By contrast, the main contribution to the interference correction comes from short ballistic  trajectories with a few scattering events~\cite{GZ,Dyakonov,DKG97}.
Experimentally, the ballistic regime can be more easily achieved in high-quality  heterostructures with high electron mobility. The point is that   in such structures, the interval of fields, where $l_B<l$ but at the same time the magnetic field is classically weak, can be very wide.

Positive magnetoconductivity due to weak localization in the ballistic regime was calculated in Refs.~\cite{GZ,DKG97}. In the presence of a moderate spin-orbit splitting of the spectrum, the ``ballistic'' magnetoconductivity was obtained in Refs.~\cite{Golub_2005,GG_FTP_2006,GlazovGolub_2008}. These results were used to fit the
weak-localization~\cite{Hamilton} and weak-antilocalization~\cite{GlazovGolub_2008,Spirito} data in various high-mobility heterostructures.

In Refs.~\cite{Golub_2005,GG_FTP_2006,GlazovGolub_2008}, the spin-orbit splitting was assumed to be comparable to or even larger than the momentum scattering rate $\hbar/\tau$, see Fig.~\ref{fig:regimes}. In this case, the spin dynamics can be well described by electron spin rotations in the effective momentum-dependent magnetic field~\cite{Lyub,GolubGanichevReview}. However, when the spin-orbit splitting becomes of the order of the Fermi energy, effects of spin-orbit interaction on the electron orbital motion can not be neglected in the calculation of the conductivity correction. 

Recently, 2D systems have become available where such an ultra-strong splitting can be realized.
Examples are electrons near the surface of polar semiconductors and at LAO/STO interfaces, or holes in HgTe-based quantum well structures with a large spin-orbit splitting~\cite{Large_R_BiTeX,LAO_STO,Minkov_2014}. In such systems, the spin energy branches are well separated (see Fig.~\ref{fig:regimes}, right panel), which results in a strongly coupled dynamics of electron spin and orbital degrees of freedom. The classical conductivity 
in such systems was analyzed in Ref.~\cite{Large_R_classical}.
Weak localization for well-separated spin branches was considered in Ref.~\cite{GDK_98,Skvortsov} in the diffusive regime and in zero magnetic field only. Recently, weak localization in spin-orbit metals based on the HgTe quantum wells has been examined in the model of double-degenerate branches of the massive  Dirac fermions~\cite{Tkachov,Richter,OGM,GKO,GKOM}.

In the present work, we develop a theory of weak localization for the systems with an arbitrary large splitting of the spin branches.
We study the quantum interference in the presence of short-range disorder potential which provides efficient inter-branch scattering.
We consider contributions to the anomalous magnetoconductivity from an arbitrary number of scatterers and derive  a general  expression for the magnetocunductivity valid in both diffusion and ballistic regimes of  weak localization.

The paper is organized as follows. In Section~\ref{Model} we formulate the model. In Section~\ref{Cond_calc} we present the derivation of the
interference-induced conductivity correction. In Section~\ref{Res_Disc}, the results for the magnetoconductivity and the zero-field correction are presented and discussed. Section~\ref{Concl} summarizes our conclusions.

\section{Model}
\label{Model}

The Hamiltonian of 2D electrons has the form
\begin{equation}
\label{Hamiltonian}
	H= \frac{\hbar^2k^2}{2m} + \alpha (\bm \sigma \times \bm k)_z,
\end{equation}
where $\bm k$ is a 2D momentum, $z$ is a normal to the structure, $m$ is the effective mass, $\bm \sigma$ is a vector of Pauli matrices, and $\alpha$ is the Rashba constant. The isotropic energy spectrum consists of two branches labeled by the index $s=\pm$:
\begin{equation}
	E_s(k) = \frac{\hbar^2k^2}{2m} + s\alpha k
\label{spectrum}
\end{equation}
with the splitting $\Delta = 2\alpha k$, Fig.~\ref{fig:regimes}.
It is worth noting that the same spectrum describes
electrons with a $\bm k$-linear isotropic 2D Dresselhaus spin-orbit interaction~\cite{GolubGanichevReview,DK}
with the substitution of $\alpha$ by the 2D Dresselhaus constant. The eigenfunctions in the two branches are spinors
\begin{equation}
\label{chiral_basis}
	|  \bm k, s \rangle =  e^{i\bm k \cdot \bm r} \frac{1}{\sqrt{2}} 
	\left( \begin{array}{cc}
	1 \\ -{\rm i}s {\rm e}^{\rm i\varphi_{\bm k}}
	\end{array}
	\right),
\end{equation}
 where $\varphi_{\bm k}$  is the polar angle of $\bm k.$

The spectrum~\eqref{spectrum} is approximately linear in the vicinity of $k=0$, where the $\pm$ bands touch each other
 (see Fig.~\ref{fig:regimes}, right panel).
In what follows, we will term this special point the ``Dirac point''.
We will first consider the situation when the Fermi energy is located above the Dirac point.
In this case, the eigenstates at the Fermi level belong to two different branches, and Fermi wavevectors $k_{\rm F}^\pm$ are different:
\begin{equation}
\label{kF}
k_{\rm F}^\pm=\frac{m}{\hbar} \left(  v_F \mp \frac{\alpha}{\hbar} \right).
\end{equation}
Here
\begin{equation}
\label{vF}
	v_{\rm F}=\sqrt{{2E_{\rm F}}/{m}+ {\alpha^2}/{\hbar^2}}
\end{equation}
is the Fermi velocity equal in both branches and $E_{\rm F}$ is the Fermi energy counted from the Dirac point.

Disorder  leads to the following types of scattering processes: intra-branch (++ and $--$) and inter-branch ($+-$ and $-+$).  In this paper we consider the short-range Gaussian disorder,
 $$ \overline{V(\bm r) V(\bm r')}=V_0^2 \delta(\bm r-\bm r').$$
Here $\overline{\cdots}$ stands for  averaging over disorder realizations,
and $V_0$ quantifies the strength of the scattering potential.
The scattering matrix element between the states $s,\bm k$ and $s',\bm k'$ is given by
\be
\langle \bm k' s'|  V| \bm k s\rangle= A_{\bm k' \bm k} V_{s's}
\ee
where 
$A_{\bm k' \bm k}=\int V(\bm r) e^{i(\bm k -\bm k')\mathbf r} d\bm r/V_0,$
$$
V_{s's}=   V_0{(1+ss'{\rm e}^{-{\rm i}\theta}) / 2},
$$
 and $\theta=\varphi_{\bm k'}-\varphi_{\bm k}$ is the scattering angle.
Importantly,  the short-range potential    provides effective inter-branch scattering for an arbitrary spin-orbit splitting.
The total (quantum) disorder-averaged scattering rate is the same in both branches:
\begin{equation}
\label{tau}
	\frac{1}{ \tau} = \frac{2\pi }{ \hbar} \left< |V_{++}|^2 g_+ + |V_{+-}|^2g_-\right>_\theta = \frac{m}{ \hbar^3}V_0^2.
\end{equation}
Here the angular brackets denote averaging over $\theta$, and the densities of states at the Fermi energy in the branches are given by
\begin{equation}
\label{g}
	g_\pm = \frac{m }{ 2\pi\hbar^2} \left(1 \mp R \right).
\end{equation}
The parameter $R$ is introduced according to
\begin{equation}
	R = \frac{\alpha }{ \hbar v_{\rm F}}.
\end{equation}

As shown in Appendix~\ref{Drude}  
(see also Ref.~\onlinecite{Large_R_classical}), the classical Drude conductivity is given by
\begin{equation}
	  \sigma_D= e^2v_{\rm F}^2\tau \frac{m}{2\pi\hbar^2}(1+R^2) 
	  = \frac{n e^2 \tau}{m},
\end{equation}
where the 2D electron concentration is 
\begin{equation}
\label{conc}
	n = \frac{m^2v_{\rm F}^2}{2\pi\hbar^2} (1+R^2).
\end{equation}

When the Fermi energy is located below the Dirac point (${R>1}$), the Fermi contour also consists of two concentric circles, ``1'' and ``2'', but they both belong to the outer spin branch $s=-$, Fig.~\ref{fig:regimes}.
The Fermi wavevectors $k_{\rm F}^{(1,2)}$ are substantially different, while the Fermi velocities in the branches are equal in this case as well.
The densities of states are given by
\begin{equation}
\label{g12}
	g^{(1,2)}=\frac{m}{2\pi\hbar^2}(R \pm 1),
\end{equation}
and the concentration is given by Eq.~\eqref{conc} as well.

\section{Conductivity calculation}
\label{Cond_calc}

The quantum correction to the conductivity in systems with spin-orbit interaction can be calculated by two approaches. The first one uses the basis of electron states with definite spin projections on the $z$ axis, $\uparrow$ and $\downarrow$. In this approach, the conductivity correction is presented as a result of interference of electronic waves with a definite total angular momentum: the interference amplitude, Cooperon, is a sum of contributions from singlet and triplet states~\cite{ILP,Golub_2005,GG_FTP_2006,GlazovGolub_2008,MN_NS_ST_SSC,MN_NS_intervalley}.
An alternative approach uses the basis of chiral states~\eqref{chiral_basis}.
This approach has been used for calculation of the conductivity correction in zero magnetic field and, recently, for calculation of its magnetic field dependence in HgTe quantum wells~\cite{GDK_98,Skvortsov,OGM,GKO,GKOM}.
In the present work we use both approaches and demonstrate that they lead to the same results. In this Section we derive the conductivity correction working in the basis of singlet and triplet states. In Appendix~\ref{AppendixB} and Supplemental Material~\cite{SM}, we derive the correction in the basis of chiral states. 

We investigate the two cases when the Fermi level is above
and below the Dirac point, Fig.~\ref{fig:regimes}.
We start with the first case, corresponding to $R<1$.

\subsection{Fermi level above the Dirac point}

The retarded (R) and advanced (A) Green functions in the subband $s$ are given by
\begin{equation}
	G^{R,A}_s (\bm r,\bm r') = G_{0}^{R,A}(\bm r,\bm r'; k_{\rm F}^s)
		\frac{1 }{ 2} \left(
		\begin{array}{cc}
	1 & \pm {\rm i} s {\rm e}^{-{\rm i}\phi} \\ \mp {\rm i} s {\rm e}^{{\rm i}\phi} &1
	\end{array}
	\right).
\end{equation}
Here $\phi$ is the polar angle of the vector $\bm \rho = \bm r-\bm r'$, $k_{\rm F}^s$ are the wavevectors at the Fermi level in two subbands, Eq.~\eqref{kF},  and
$G_{0}^{R,A}(\bm r,\bm r'; k_{\rm F})$ is the standard Green function in a simple parabolic band with the Fermi wavevector $k_{\rm F}$:
\begin{align}
	G_{0}^{R,A}(\bm r,\bm r'; k_{\rm F}) & =  \mp {\rm i}  
	\frac{\sqrt{k_{\rm F}}}{ \hbar v_{\rm F} \sqrt{2\pi \rho}} \\
	&\times \exp{\left[\pm k_{\rm F}\rho-\frac{\rho}{ 2l} \mp {\rm i}\frac{\pi}{ 4} +\frac{{\rm i}}{2}\Phi(\bm r,\bm r') \right]} \nonumber 
\end{align}  
with $l=v_{\rm F}\tau$. The magnetic field induced phase is
\begin{equation}
\label{Phi_phase}
	{\Phi(\bm r,\bm r') = (y+y')(x'-x)/ l_B^2},
\end{equation}
 where
\[l_B = \sqrt{\hbar/|eB|}\]
is the magnetic length for elementary charge ($e<0$).
Here we used the fact that $\tau$ and $l$ are the same in both branches.

\begin{figure}[t]
\includegraphics[width=\linewidth]{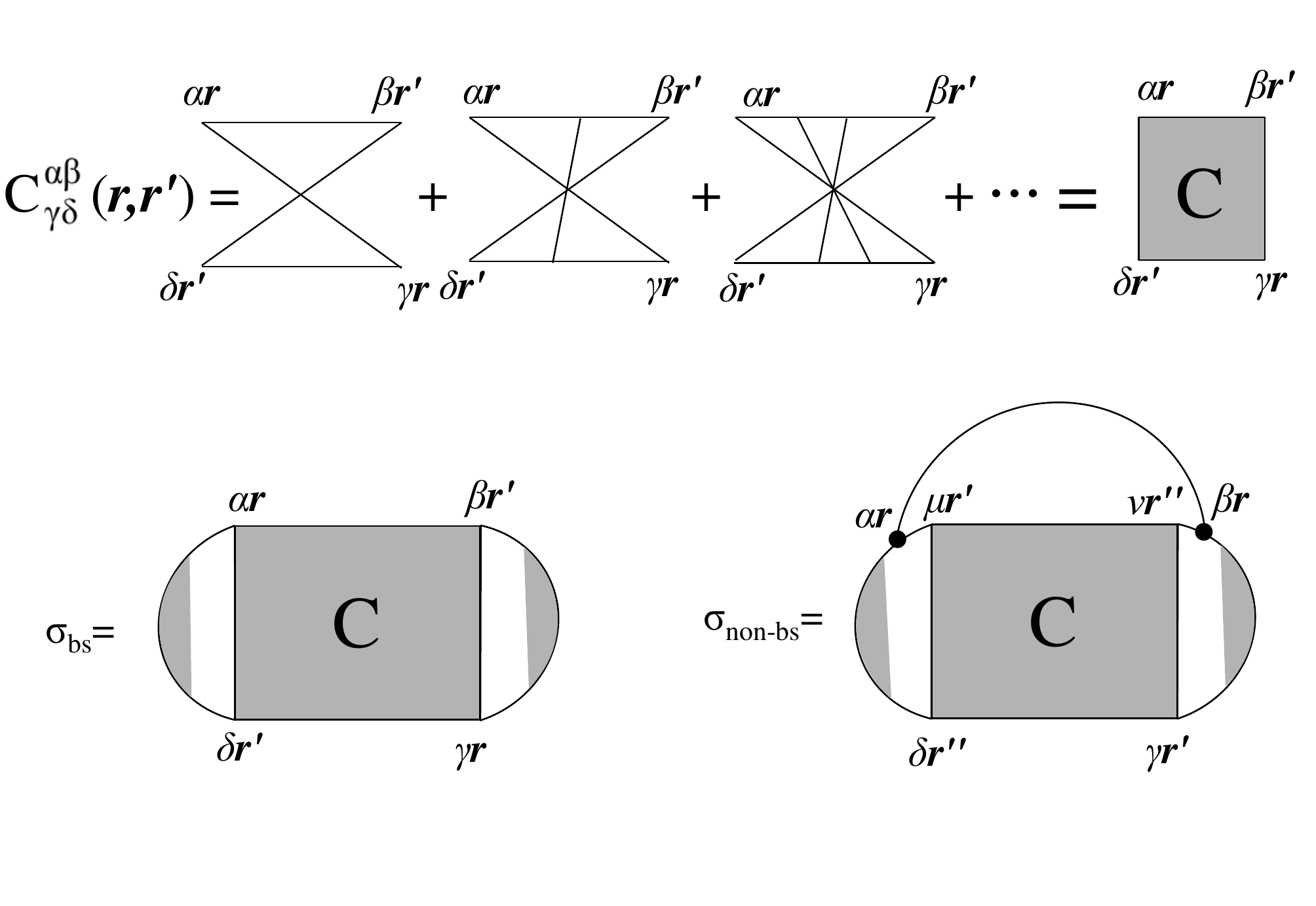}
\caption{Diagrammatic Cooperon equation and conductivity corrections.}
\label{fig:diagrammes}
\end{figure}

The interference-induced correction to the conductivity is expressed via the Cooperon, see Fig.~\ref{fig:diagrammes}.
In the basis of states with spin projections on the $z$ axis,
${\alpha, \beta, \gamma, \delta = \uparrow, \downarrow}$, the Cooperon satisfies the following equation:
\begin{equation}
\label{C2}
	{\cal C}^{\alpha\beta}_{\gamma\delta}(\bm r_1,\bm r_2)= V_0^2 P^{\alpha\beta}_{\gamma\delta}(\bm r_1,\bm r_2)
	+\int\! d \bm r P^{\alpha\mu}_{\gamma\nu}(\bm r_1,\bm r)  {\cal C}^{\mu\beta}_{\nu\delta}(\bm r,\bm r_2),
\end{equation}
where 
$$P^{\alpha\mu}_{\gamma\nu}(\bm r_1,\bm r) = V_0^2 G^R_{\alpha\mu}(\bm r_1,\bm r) G^A_{\gamma\nu}(\bm r_1,\bm r),$$  and  $G^{R,A}_{\alpha\mu}=\sum\limits_s \left(G^{R,A}_s \right)_{\alpha\mu}$.

Under the condition of well-separated spin branches, $|k_{\rm F}^+-k_{\rm F}^-|l \gg 1$, which we assume from now on,
the product $G^R_s(\bm r_1,\bm r)G^A_{s'}(\bm r_1,\bm r)$ oscillates rapidly
on the scale of the mean-free path if $s \neq s'$.
Therefore $\hat{P}$ has only two terms in the sum with $s=s'$:
\begin{equation}
	P^{\alpha\mu}_{\gamma\nu}(\bm r_1,\bm r) = V_0^2 \sum_{s=\pm}  (G^R_s)_{\alpha\mu}  (G^A_{s})_{\gamma\nu}.
\label{Pam}
\end{equation}
Summation over $s$ yields the 16 components of the matrix $\hat{P}$.
Passing to the basis of states with a fixed total angular momentum
and its projection to the $z$ axis,
i.e. to the basis $\uparrow \uparrow, (\uparrow \downarrow + \downarrow\uparrow)/\sqrt{2}, (\uparrow \downarrow - \downarrow\uparrow)/\sqrt{2}, \downarrow\downarrow$ we obtain
that the triplet state with the zero momentum projection, i.e. ${(\uparrow \downarrow + \downarrow\uparrow)/\sqrt{2}}$,
does not contribute to weak localization.
The matrix $\hat{P}$
corresponding to the three other states, ${\uparrow \uparrow, (\uparrow \downarrow - \downarrow\uparrow)/\sqrt{2}, \downarrow\downarrow}$,
has the following form:
\begin{equation}
\label{P4}
 \hat{P} = P_0 \left(
		\begin{array}{ccc}
	\frac{1}{ 2} &  -{\rm i} \frac{R }{ \sqrt{2}} {\rm e}^{-{\rm i}\phi} & \frac{1}{ 2}{\rm e}^{-2{\rm i}\phi}\\
	{\rm i} \frac{R }{ \sqrt{2}} {\rm e}^{{\rm i}\phi}&1 & {\rm i} \frac{R }{ \sqrt{2}} {\rm e}^{-{\rm i}\phi}  \\
	\frac{1}{ 2}{\rm e}^{2{\rm i}\phi}& -{\rm i} \frac{R }{ \sqrt{2}} {\rm e}^{-{\rm i}\phi}&\frac{1}{ 2}
	\end{array}
	\right),
\end{equation}
where
\begin{equation}
	P_0(\bm r,\bm r') = \frac{\exp{(-|\bm r-\bm r'|/\tilde{l})} }{ 2\pi l |\bm r-\bm r'|} {\rm e}^{{\rm i}\Phi(\bm r,\bm r')}.
	\label{P0r}
\end{equation}
with $\tilde{l}=l/(1+\tau/\tau_\phi)$, and $\Phi(\bm r,\bm r')$ given by Eq.~\eqref{Phi_phase}.

It is worth comparing the form of the matrix $\hat{P}$, Eq.~\eqref{P4}, with its form for weakly split Fermi circles at $\Delta \ll E_{\rm F}$. In that case,  the matrix $\hat{P}$ had the form of a $4\times 4$ block-diagonal matrix with a separate triplet $3\times 3$ block and an independent singlet sector. In the limit $\Delta\tau/\hbar \to \infty$ (but still $R=0$) the triplet state with zero spin projection decouples and its matrix elements vanish, so that the triplet block becomes a $2\times 2$ matrix~\cite{GG_FTP_2006}. However, Eq.~\eqref{P4} demonstrates that, for strongly split 
spin branches ($R\neq 0$), the singlet Cooperon state becomes mixed with the two triplet ones. This mixing, linear in the parameter $R$, arises due to the difference in the densities of states in the spin branches: ${g_+ - g_- \propto R}$, Eq.~\eqref{g}.

The Cooperon can be found in the basis of Landau level states with charge $2e$, $\Psi_{Nq}(\bm r)$:
$$
{\cal C}^{\alpha\beta}_{\gamma\delta}(\bm r,\bm r')=\sum_{N,N',q,q'} {\cal C}^{\alpha\beta}_{\gamma\delta}(N,N')
\Psi^*_{Nq}(\bm r) \Psi_{N'q'}(\bm r'),$$
where $N=0,1,2\ldots$ is the Landau level number, and $q$ is the in-plane wavevector for the Landau gauge.
Expanding the matrix $\hat{P}$, Eq.~\eqref{P4}, over this basis, we obtain from Eq.~\eqref{C2} an
infinite system of linear equations for the coefficients ${\cal C}^{\alpha\beta}_{\gamma\delta}(N,N')$. It  can be block-diagonalized in the basis of states with fixed $N+s_z$,
where $s_z=1,-1$ is the angular momentum projection in the triplet state while $s_z=0$ describes the singlet. 
The equation {for the blocks ${\cal C}(N)$ }has the following form:
\begin{equation}
	{\cal C}(N) = V_0^2A_N + A_N{\cal C}(N),
	\end{equation}
where	
	\begin{equation}
	\label{P_N}	
	A_N = \left( \begin{array}{ccc}
	\frac{1}{2}P_{N-2}  & {\rm i} \frac{R }{ \sqrt{2}} P_{N-2}^{(1)} & \frac{1}{2}P_{N-2}^{(2)} \\
	{\rm i} \frac{ R }{ \sqrt{2}} P_{N-2}^{(1)} & P_{N-1} & -{\rm i} \frac{R }{ \sqrt{2}} P_{N-1}^{(1)}\\
	\frac{1}{2}P_{N-2}^{(2)} &  -{\rm i}\frac{R }{ \sqrt{2}} P_{N-1}^{(1)} &  \frac{1}{2}P_{N}
	\end{array} \right).
\end{equation}
Here $P_N^{(m)}$ ($m=1,2$) and $P_N \equiv P_N^{(0)}$ are defined as follows:
\begin{align}
	P_N^{(m)} = &\frac{l_B }{ l} \sqrt{\frac{N!}{ (N+m)!}} \\
	&\times \int\limits_0^\infty dx \exp\left(-x {l_B /\tilde{l}} - {x^2/ 2} \right) x^m L_N^m(x^2) \nonumber
\end{align}
with $L_N^m$ being the Laguerre polynomials. All values with negative indexes should be substituted by zeros.

The conductivity correction is a sum of two contributions shown in Fig.~\ref{fig:diagrammes}:
\[
\sigma = \sigma_\text{bs} + \sigma_\text{non-bs}.
\]
The backscattering contribution to the magnetoconductivity is given by
\begin{equation}
	\sigma_\text{bs} = \frac{\hbar }{ 4\pi} \int d \bm r \int d \bm r' \sum_{\alpha\beta}\left[ \tilde{\cal C}(\bm r,\bm r') \Gamma(\bm r,\bm r') \right]^{\alpha\beta}_{\beta\alpha},
\end{equation}
where $\tilde{\cal C}={\cal C}-V_0^2P$ is the Cooperon calculated starting from three scattering lines. The squared electric current vertex is presented by the following operator
\begin{equation}
	{\Gamma} = \sum_{s=\pm} \left( \frac{{\rm i} e l }{ \hbar}\frac {\tau_{\rm tr}^{(s)} }{ \tau}\right)^2 G^R_s G^A_s,
\end{equation}
where $\tau_{\rm tr}^{(s)}$ is the transport time in the subband $s$.
Comparing with Eq.~\eqref{Pam}, we see that  operator $\Gamma$  has a matrix form similar to $P$, the only modification is caused by the squared transport time.

In contrast to the quantum scattering rates, the transport rates in the branches are different. In order to calculate the transport times, we solve a system of equations for the velocity vertexes in the subbands, $v_s^x$:
\begin{equation}
	v_s^x(\varphi) = v_{\rm F} \cos{\varphi} + \sum_{s'=\pm} (1-s'R) \left<
	\frac{1+ss'\cos{\theta} }{ 2}  v_{s'}^x(\varphi') \right>_{\varphi'}.
\end{equation}
The solution is given by 
$$v_s^x(\varphi)=v_{\rm F} \cos{\varphi} \: \tau_{\rm tr}^{s}/\tau,$$ 
where~\cite{Large_R_classical}
\begin{equation}
	\frac{\tau_{\rm tr}^{\pm}}{ \tau}= 1 \mp R.
\end{equation}

As a result, we obtain:\footnote{Strictly speaking, formally we get 
${\rm Tr} \left[\Pi \bar{A}_N A_N^2 \left({\cal I}-A_N\right)^{-1}\right]$ where $\bar{A}_N$ is obtained from the matrix $A_N$ by the substitution 
${R \to R(3+R^2)/(1+3R^2)}$. However, replacing the matrix 
$\bar{A}_N$ by $A_N$ does not change the trace.}
\begin{align}
\label{sigma_a_fin}
	\sigma_\text{bs}(B) = - \frac{e^2 }{ 2\pi^2 \hbar}  & (1+3R^2) \left( \frac{l }{ l_B} \right)^2 \\
	& \times \sum_{N=0}^\infty {\rm Tr} \left[\Pi A_N^3 \left({\cal I}-A_N\right)^{-1}\right], \nonumber
\end{align}
where ${\cal I}$ is a $3\times 3$ unit matrix, and
\begin{equation}
	\Pi = \text{diag}(1,-1,1).
\end{equation}

The non-backscattering contribution, given by a sum of the second diagram in Fig.~\ref{fig:diagrammes} and the one conjugated to it, reads:
\begin{align}
	&\sigma_\text{non-bs} = \frac{\hbar }{ \pi} \\
	& \times \sum_{\alpha\beta} \int d \bm r \int d \bm r' \int d \bm r''
	\left[ K(\bm r, \bm r') {\cal C}(\bm r',\bm r'') K(\bm r'',\bm r) \right]^{\alpha\beta}_{\beta\alpha}. \nonumber
\end{align}
Here the vertex is given by
\begin{equation}
	K(\bm r, \bm r')=\frac{{\rm i} e l }{ \hbar}\! \sum_{s=\pm}\!    \frac{\tau_{\rm tr}^{(s)} }{ \tau} V_0^2 G^R_s(\bm r, \bm r') G^A_s(\bm r, \bm r') \cos(\varphi-\varphi'),
\end{equation}
where $\varphi$ and $\varphi'$ are polar angles of the vectors $\bm r$ and $\bm r'$, respectively.
Summation over $s$ yields
\begin{align}
	K(\bm r,\bm r') =  \frac{{\rm i} e l }{ \hbar} \cos{(\varphi-\varphi')} (1+R^2) P({R'}),
\end{align}
where the matrix $P(R')$ is given by $P$, Eq.~\eqref{P4}, with $R$ replaced by
\begin{equation}
R'=\frac{2R }{ 1+R^2}.
\end{equation}
In the basis  of Landau level states with charge $2e$ and fixed $N+s_z$,
the operator $K(\bm r,\bm r')$ is written as
\begin{equation}
	K = \frac{1}{ 2}K_N^T - \frac{1}{ 2}K_{N+1},
\end{equation}
where the matrix $K_N$ is
\begin{equation}
\label{K_N}
	K_N = \left( \begin{array}{ccc}
	\frac{1}{ 2}P_{N-2}^{(1)}  & -{\rm i} \frac{R'}{ \sqrt{2}} P_{N-2}^{(2)} & \frac{1}{ 2}P_{N-2}^{(3)} \\
	{\rm i} \frac{R'}{ \sqrt{2}} P_{N-1} & P_{N-1}^{(1)} & {\rm i} \frac{R'}{ \sqrt{2}} P_{N-1}^{(2)}\\
	-\frac{1}{ 2}P_{N-1}^{(1)} &  -{\rm i}\frac{R'}{ \sqrt{2}} P_N &  \frac{1}{ 2}P_{N}^{(1)}
	\end{array} \right).
\end{equation}
Finally,
we obtain
\begin{align}
\label{sigma_b_fin}
	\sigma_\text{non-bs}(B) = & \frac{e^2 }{ 4\pi^2 \hbar} \left( 1+R^2 \right)^2 \left( \frac{l }{ l_B} \right)^2  \\
	 \times & \sum_{N=0}^\infty {\rm Tr} \Biggl[K_N \Pi K_N^T A_N \left({\cal I}-A_N\right)^{-1} \nonumber
	\\ & + K_N^T \Pi K_N A_{N+1} \left({\cal I}-A_{N+1}\right)^{-1}\Biggr].  \nonumber
\end{align}

\subsection{Fermi level below the Dirac point}
\label{Below_DP}

For the Fermi level below the Dirac point, $R>1$ in Fig.~\ref{fig:regimes},
the Green functions for the two Fermi circles $i=1,2$ are different due to unequal values of the Fermi wavevectors $k_{\rm F}^{(1,2)}$.
Therefore we have
\begin{equation}
	P(\bm r, \bm r')=V_0^2 \sum_{i=1,2}  G^R_i(\bm r, \bm r')  G^A_i(\bm r, \bm r').
\end{equation}
The products $G^R_{i}G^A_{i}$ have the same coordinate dependence as at the Fermi level above the Dirac point;
the difference is only in the density of states factors
$g_i$, Eq.~\eqref{g12}.
As a result, $P(\bm r, \bm r')$ has the same form as in a one-subband system with $s=-$ with
the density of states equal to $g_1+g_2$.
The conductivity correction in such a system is the same as in the single branch $s=-$ with $R=1$ and the transport time $\tau_{\rm tr}=2\tau$.
Therefore, the corrections $\sigma_\text{bs}$ and $\sigma_\text{non-bs}$ are given by Eqs.~\eqref{sigma_a_fin},~\eqref{sigma_b_fin} with $R=1$.

The above consideration shows that the conductivity correction at $R>1$ is equal to that at $R=1$. In other words, when the Fermi level goes down through the Dirac point at $k=0$, the correction does not change with further decreasing the Fermi energy to the bottom of the conduction band.

\section{Results and Discussion}
\label{Res_Disc}

Let us now discuss the obtained expressions for the conductivity corrections.
We remind the reader that Eqs.~(\ref{sigma_a_fin}) and~(\ref{sigma_b_fin}) have been obtained
under the condition $|k_{\rm F}^+-k_{\rm F}^-|l \gg 1$.
At small Rashba splitting relative to the Fermi energy, $R \to 0$,
the derived conductivity corrections coincide
with the result obtained in Refs.~\cite{Golub_2005,GlazovGolub_2008} for weakly split spin subbands, 
$|k_{\rm F}^+-k_{\rm F}^-|l \ll 1$ in the limit of fast spin rotations $\Delta \, \tau/\hbar \to \infty$.

At the Fermi level lying exactly in the Dirac point, $R =1$,
our results pass into the expressions obtained in Ref.~\cite{MN_NS_ST_SSC}
for the spectrum consisting of a single massless Dirac cone.  
A single-cone result for the considered two-subband system at $R=1$ follows from the zero density of states of the  subband $s=+$ (the Fermi wavevector for it is equal to zero). In this case, the contribution of this subband to the conductivity is zero and  scattering to it from the other branch is absent. Therefore the subband $s=+$ is excluded from transport while the states in the other branch ($s=-$) are described by the same spinors as in a single valley of graphene or on the surface of a three-dimensional topological insulator. The same result is obtained for a single spin subband in a 2D topological insulator at the critical quantum-well width 
(no gap) in Ref.~\cite{GKO} (this corresponds to $\eta=1$ there).

\subsection{Zero field conductivity correction}

At zero field, the conductivity corrections are obtained from Eqs.~\eqref{sigma_a_fin},~\eqref{sigma_b_fin} by passing to integration over $N$.
This yields:
\begin{widetext}
\begin{align}
\label{sigma_a_zero_field}
	\sigma_\text{bs}(B\!=\!0)&\!=\!-\frac{e^2 (1+3R^2) }{ 4\pi^2 \hbar} 
	\!\int\limits_0^{1/(1+\gamma)}\!\frac{dP}{P^3} 
	{\rm Tr} \left[\Pi A^3 \left({\cal I}-A\right)^{-1}\right],
	\\
\label{sigma_b_zero_field}
	\sigma_\text{non-bs}(B\!=\!0)&\!= \frac{e^2 }{ 8\pi^2 \hbar} 
	\left( 1+R^2 \right)^2  
\int\limits_0^{1/(1+\gamma)} \frac{dP}{P^3}  
{\rm Tr} \left[(K\Pi K^T+K^T\Pi K)A 
\left({\cal I}-A\right)^{-1}\right]. 
\end{align}
Here the matrices $A$ and $K$ are obtained from $A_N$ and $K_N$ ($N \gg 1$) by substitutions
~\cite{GKO,MN_NS_intervalley}:
\begin{equation}
	P_N \approx \frac{1 }{ \sqrt{4N(l/l_B)^2+(1+\gamma)^2}} \equiv P,
	\qquad 
	P_N^{(m)} \approx P \left[\frac {1-P(1+\gamma)}{1+P(1+\gamma)} \right]^{m/2},
	\label{PNm_zero_field}
		\end{equation}	
		with
$\gamma = {\tau/\tau_\phi}$.
Note that $P$ is expressed~\cite{SM} as the Fourier transform with respect to 
$\bm r-\bm r'$ of the function $P_0(\bm r,\bm r')$ defined in Eq.~(\ref{P0r}) with 
$\Phi(\bm r,\bm r')=0$. The corresponding wave vector is given by $q=2N/l_B$.

Calculating the traces in Eqs.~\eqref{sigma_a_zero_field} and\eqref{sigma_b_zero_field} we get:
\begin{equation}
\label{sigma0_tot}
	\sigma(B=0)  = -\frac{e^2 }{\pi^2 \hbar}\int\limits_0^{1/(1+\gamma)} dP \:
	{\frac {{\cal K}(P,R,\gamma)}{ \left[
	1- \left( 1+\gamma \right)  \left( {R}^{2}-1 \right) {P}^{3}
	- \left(2 \,\gamma+2  -{R}^{2} \right) {P}^{2}+P\gamma \right]  \left( P\gamma+1 \right)
 \left( 1+P+P\gamma \right) ^{3}}}.
\end{equation}
\end{widetext}
The explicit form of the function ${\cal K}(P,R,\gamma)$ is derived in Supplemental Material~\cite{SM}.
When one is interested in the corrections up to $\mathcal{O}(1)$ in the
limit $\gamma\to 0$, $\gamma$ in the numerator of  Eq.~(\ref{sigma0_tot}) can be neglected, which yields
\begin{align}
& {\cal K}(P,R,0)=	\left( R^2-1 \right)  \left( 3\,{R}^{4}+16\,{R}^{2}
+9 \right) {P}^{4}
\nonumber
\\
& + \left( 10-5\,{R}^{6}+17\,{R}^{2}+6\,{R}^{4} \right){P}^{3} \nonumber \\
&+ \left(
14+11\,{R}^{2}-6\,{R}^{4}+{R}^{6} \right){P}^{2}
\nonumber \\
& + \left( 2-10\,{R}^{4}-5\,{
R}^{2}+{R}^{6} \right) P
-(1+3\,{R}^{4}).
\end{align}
Furthermore, one can also neglect $\gamma$ in the non-singular factors in the denominator.
For this reason, finding the roots of the cubic polynomial in the square brackets in the denominator of Eq.~(\ref{sigma0_tot}) perturbatively in $\gamma\ll 1$, we replace it by
\begin{equation}
	[1+P-P^2(1-R^2)] \left( 1-P - \gamma \frac{R^2}{1+R^2}\right).
\end{equation}
Then, integration in Eq.~(\ref{sigma0_tot}) yields
\begin{equation}
\label{sigma0_gen}
	\sigma(B\!=\!0)= \sigma_{\text{diff}} (T) + \sigma_\text{ball}(R).
\end{equation}
The first term here is the diffusive contribution dependent on the temperature $T$ via $\tau_\phi$:
\begin{equation}
\label{diff_zero_field}
	\sigma_{\text{diff}} (T)	= \frac{e^2 }{ 4\pi^2 \hbar} \ln{\left[ \frac{\tau_\phi(T)}{\tau}\right]}.
\end{equation}
We emphasize that the coefficient in front of the logarithm  is independent of $R.$  It is worth stressing that  the coefficients in  divergent logarithmic terms  related to both  backscattering and non-backscattering contributions are $R-$dependent~\cite{SM}, and only the sum of these terms gives the universal coefficient   ${e^2 }/{(4\pi^2 \hbar)}$ prescribed by  symplectic  class of symmetry.

It is shown in Appendix~\ref{dephasing} that, in the experimentally relevant case of fixed electron concentration, the dephasing time $\tau_\phi$ is independent of $R,$ so that the argument of the logarithm in Eq.~\eqref{diff_zero_field} also does not depend on $R$.
The dependence on $R,$ however, appears in the ballistic term which takes into account the interference corrections from ballistic trajectories with a few (three or more, because we discuss the contribution sensitive to magnetic field) scattering events and is regular at low temperatures 
$\tau/\tau_\phi\ll 1.$
At $\tau/\tau_\phi=0$ we obtain the following analytical expression for the ballistic contribution to the conductivity correction:
\begin{align}
\sigma_\text{ball}(R) &=
-\frac{e^2}{4\pi^2 \hbar}
\left\{\frac{3+8R^2+R^4}{4} \right.
\nonumber
\\
&\!\!\! - \frac{4-R^2(1+R^2)}{2}\ln 2 -
\frac{2+R^2+R^4}{4}\ln(1+R^2) \nonumber
\\
&\!\!\! + \left. \frac{8-13R^2+5R^4}{2\sqrt{5-4R^2}}
\ln\frac{3+\sqrt{5-4R^2}}{2\sqrt{1+R^2}}\right\}.
\label{sigma-ball-tauphi-R}
\end{align}
In particular, we find 
\begin{equation}
\sigma_\text{ball}(R)=-\frac{e^2}{4\pi^2 \hbar}\left\{
                                                        \begin{array}{ll}
                                                          3/4-2\ln2+\frac{4}{\sqrt{5}}\ln\frac{3+\sqrt{5}}{2}, & R=0; \\
                                                          3-2\ln 2, & R=1,
                                                        \end{array}
                                                      \right.
                                                      \label{total-R0R1}
\end{equation}
which gives the values $-0.0275 e^2/\hbar$ at $R=0$ and $-0.041 e^2/\hbar$ at $R=1$.

\begin{figure}[ht]
\includegraphics[width=0.9\linewidth]{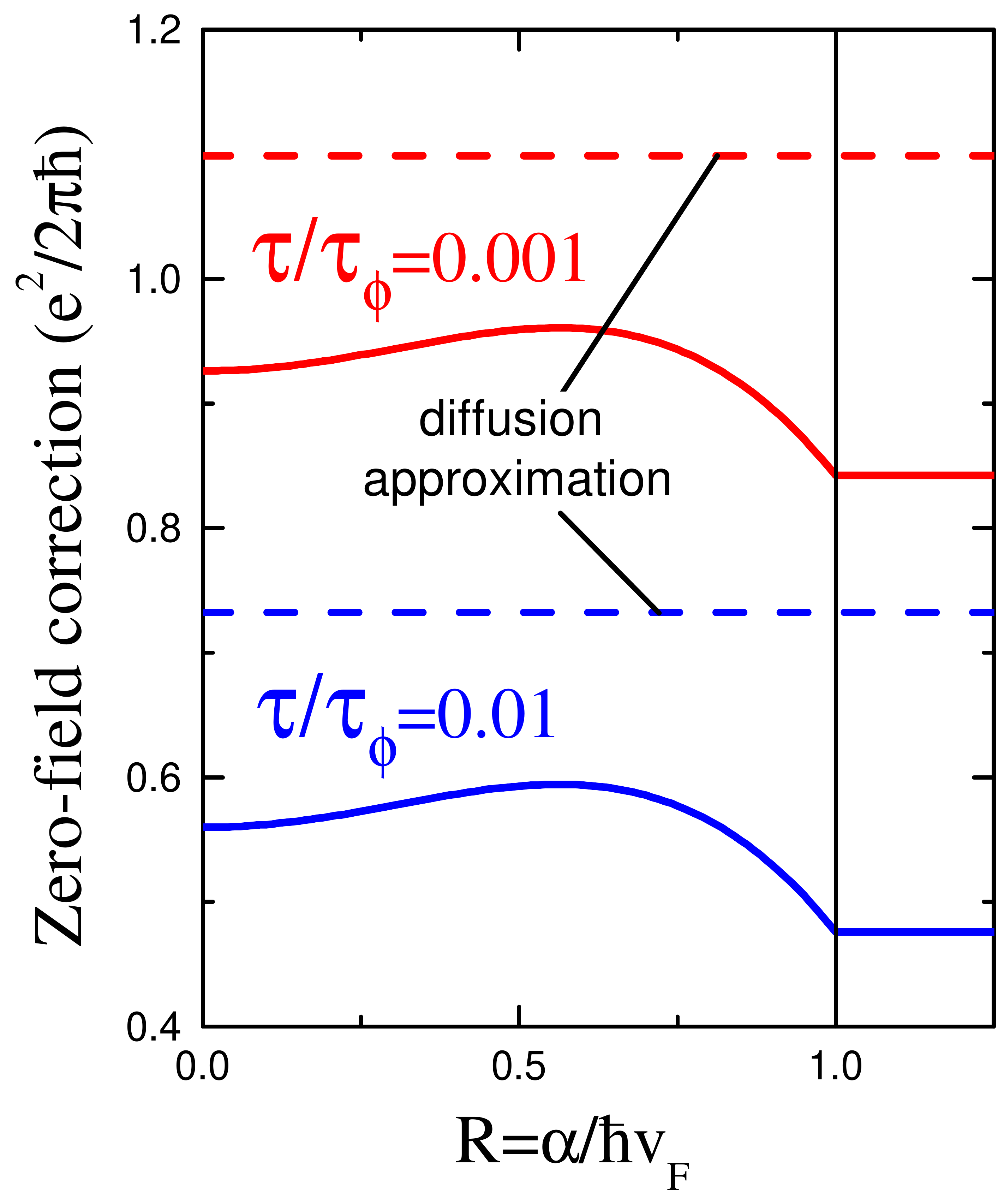}
\caption{Zero field conductivity correction as a function of the Rashba splitting parameter $R=\alpha/(\hbar v_{\rm F})$ (solid curves).
The diffusive contribution to the quantum correction, Eq.~\eqref{diff_zero_field}, is independent of $R$ (dashed curves).
}
\label{fig:zero_field}
\end{figure}

The correction $\sigma_\text{ball}(R)$ determines the dependence of the total conductivity correction on the spin-orbit splitting.
The zero-field conductivity correction is presented in Fig.~\ref{fig:zero_field}.
In contrast to the result of diffusion approximation, the total correction depends, however, on the Fermi level position at $0<R<1$. The correction  saturates at a certain value after the Fermi level crosses the Dirac point, at $R\geq 1$, see Sec.~\ref{Below_DP}.
Significant difference between the exact result and the result obtained within the diffusive approximation, Eq.~\eqref{diff_zero_field}, which is clearly seen in Fig.~\ref{fig:zero_field}, demonstrates the essential role of ballistic processes in weak localization at realistic values of $\tau/\tau_\phi$.

For a system of a finite size $L$ the conductivity correction is finite even in the absence of dephasing because the particle trajectories can not be longer than $L$. At $\gamma=0$ we integrate in Eq.~\eqref{sigma0_tot} up to ${P=1-l^2/(2L^2)}$ and obtain:
\begin{equation}
	\sigma_{L}(R) =
\frac{e^2}{4\pi^2 \hbar}\ln\frac{2L^2}{l^2}- \frac{e^2}{4\pi^2 \hbar}\ln(1+R^2) + \sigma_\text{ball}(R).
\end{equation}
This equation shows that the ballistic correction calculated without dephasing, when the diffusive contribution is cut off by the system size, differs from the result calculated at finite dephasing by the term $\propto \ln (1+R^2)$. This is related to the fact that $\tau_\phi/\tau$ corresponds to $L^2/({\rm D}\tau)$  in the logarithmic diffusive contribution, where ${\rm D}=(1+R^2)l^2/(2\tau)$ is the diffusion coefficient (see Appendix~\ref{Drude}).

As discussed above, the zero-field correction can be also obtained by calculations in the momentum space in the basis of chiral subband states~\eqref{chiral_basis}. 
This alternative derivation, leading to the same results, is presented in Appendix~\ref{AppendixB} and Supplemental Material \cite{SM}. Backscattering and non-backscattering contributions to the conductivity correction are calculated and their dependences on $R$ and on dephasing rate are analyzed. It is shown in Supplemental Material \cite{SM} that the backscattering and the non-backscattering contributions to the conductivity have the same order of magnitude and different signs compensating each other to a large extent.

\subsection{Magnetoconductivity}

\begin{figure*}[ht]
\includegraphics[width=0.8\linewidth]{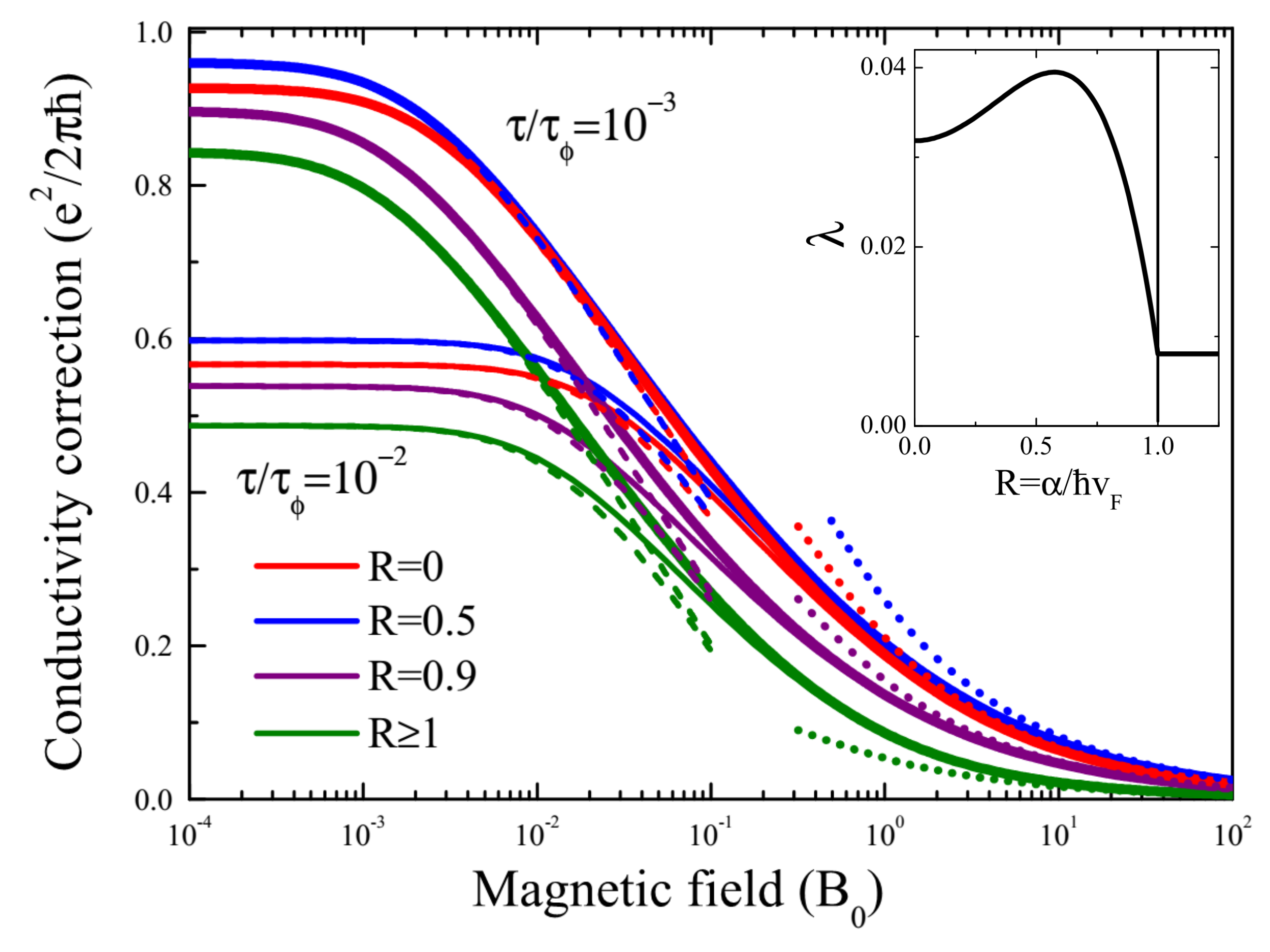}
\caption{Magnetoconductivity at different $R=\alpha/(\hbar v_{\rm F})$. The dephasing rate $\tau/\tau_\phi=10^{-2}$ (thin solid curves) and $\tau/\tau_\phi=10^{-3}$ (thick solid curves). Dashed lines present the diffusion approximation results, Eq.~\eqref{diff}. Dotted lines are high-field asymptotics Eq.~\eqref{asympt}. Inset shows the function $\lambda(R)$, Eq.~\eqref{lambda},  describing the high-field asymptotics of the conductivity correction.
}
\label{fig:mcond}
\end{figure*}

The magnetic-field dependence of the conductivity correction is given by Eqs.~\eqref{sigma_a_fin},~\eqref{sigma_b_fin}. The results of calculations are shown in Fig.~\ref{fig:mcond}. The magnetic field is given in units of the characteristic field $B_0$
\begin{equation}
	B_0 = \frac{\hbar}{ |e| l^2}.
\end{equation}
The correction monotonously decreases with the magnetic field
 for all $R$. The magnetoconductivity varies with the Rashba splitting. The magnetic field dependence is stronger at $R \approx 0.5$ due to higher zero-field value of the correction, see Fig.~\ref{fig:zero_field}.
The magnetic field dependencies at a given $R$ coincide in fields $B \gtrsim  0.1 B_0$ independently of the value of $\tau_\phi/\tau$. The reason is that the  magnetic field induced phase which breaks the interference is stronger in those fields than dephasing.

In high magnetic fields $B \gg B_0$ (but still $B$ is classically weak), the conductivity correction decreases according to the following asymptotic law:
\begin{equation}
\label{asympt}
\sigma_\text{hf} =  \frac{e^2 }{ \hbar} \lambda(R) \sqrt{\frac{B_0}{ B}}.	
\end{equation}
The function $\lambda(R)$ is a sum of backscattering and non-backscattering contributions, $\lambda = \lambda_\text{bs} + \lambda_\text{non-bs}$, where
\begin{equation}
\label{lambda_bs}
\lambda_\text{bs} = -\frac{1+3R^2}{2\pi^2} \sum_{N=0}^\infty {\rm Tr} \left(\Pi \tilde{A}_N^3 \right),
\end{equation}
\begin{align}
\lambda_\text{non-bs} &= \frac{(1+R^2)^2}{ 4\pi^2} \\
& \times \sum_{N=0}^\infty
{\rm Tr} \left(\tilde{K}_N \Pi \tilde{K}_N^T \tilde{A}_N + \tilde{K}_N^T \Pi \tilde{K}_N \tilde{A}_{N+1} \right). \nonumber
\end{align}
The matrices $\tilde{A}_N$ and $\tilde{K}_N$ are obtained from $A_N$ and $K_N$ by the following substitutions~\cite{GKO}: all $P_N^{(m)}$ with odd $N$ should be taken by zeros, and at even $N=2k$
\begin{align}
\label{P_asympt}
&		P_{2k} \to \sqrt{\frac{\pi }{ 2}}\frac{(2k)!}{ 2^{2k}(k!)^2},
		\quad
		P_{2k}^{(2)} \to \sqrt{\frac{\pi }{ 2}}\frac{\sqrt{(2k)!(2k+2)!}}{ 2^{2k+1} k! (k+1)!}, \\
&		P_{2k}^{(1)} \to -\frac{1}{ \sqrt{2k+1} },\quad
		P_{2k}^{(3)} \to -\sqrt{\frac{2k+2}{ (2k+1)(2k+3)} }.
		\nonumber
\end{align}
From Eqs.~\eqref{lambda_bs}-\eqref{P_asympt} we obtain: 
\begin{equation}
\label{lambda}
	\lambda(R) = \frac{\sqrt{\pi}(5R^2+3)}{32\sqrt{2} \Gamma^4(3/4)} - \frac{45R^4+26R^2+13}{ 256 \sqrt{2\pi}}.
\end{equation}

The dependence $\lambda(R)$ is shown in the inset to Fig.~\ref{fig:mcond}. The high-field asymptotes Eq.~\eqref{asympt} are presented in Fig.~\ref{fig:mcond} by dotted lines. 

Comparison with the results of exact calculation shows that the
asymptotes perfectly describe the conductivity correction at $B\gtrsim 10 B_0$,
but only the exact expressions describe the magnetoconductivity in the intermediate range of fields. 

The magnetoconductivity $\delta\sigma (B)	= \sigma(B)	-\sigma(0)$ in the diffusion approximation is given by the
conventional expression for systems with fast spin relaxation:
\begin{align}
\label{diff}
	\delta\sigma_\text{diff} (B)	=  -\frac{e^2 }{ 4\pi^2 \hbar}
	[\psi(1/2+1/b) + \ln{b}],
	\\
	b=  2\frac{B}{ B_0} \frac{\tau_\phi}{\tau}(1+R^2) 
	= \frac{4{\rm D}\tau_\phi}{l_B^2}, \nonumber
\end{align}
where $\psi(x)$ is digamma function.
The dependences $\sigma(0)+\delta\sigma_\text{diff} (B)$ with $\sigma(0)$ calculated by Eqs.~\eqref{sigma0_gen}-\eqref{sigma-ball-tauphi-R} are plotted in Fig.~\ref{fig:mcond} by dashed lines.
It is well known that the diffusion approximation does not describe the magnetoconductivity in fields
${B \agt B_0}$~\cite{Golub_2005,MN_NS_ST_SSC}.
Our calculation demonstrates that diffusion approximation satisfactorily describes the magnetic field dependence of the conductivity correction up to $(0.02 \ldots 0.03) B_0$, see Fig.~\ref{fig:mcond}.

Figure~\ref{fig:mcond} demonstrates that neither diffusion approximation nor high-field asymptotics describe the conductivity correction, and the exact expressions are needed to describe the magnetic field dependence in the whole range of classically-weak fields.

\section{Conclusion}
\label{Concl}

In this work we have developed the theory of weak localization in 2D systems with an arbitrary strong linear-in-$\bm k$
spin-orbit  splitting of the energy spectrum. The theory describes weak antilocalization in systems with the Rashba or
Dresselhaus isotropic spin-orbit splittings. We have
derived analytical expression for the conductivity correction that includes both diffusive and ballistic contributions and  is valid in a wide interval of the phase breaking rates and magnetic fields. We have found that the ballistic contribution depends solely on the spectrum characteristics and, therefore, reflects intrinsic properties of the system. 
We have also shown that the magnetoconductivity varies with the Fermi energy when the Fermi level is above the ``Dirac point'' of the spectrum, but 
does not depend on the Fermi energy when it goes below this point.

\section*{Acknowledgments}
We thank M.~O.~Nestoklon for helpful discussions.
Partial financial  support from the Russian Foundation for
Basic Research, the EU Network FP7-PEOPLE-2013-IRSES Grant No 612624 ``InterNoM'', by Programmes of RAS,  and  DFG-SPP1666 is gratefully acknowledged.

\appendix

\section{Drude conductivity}
\label{Drude}

Although the Drude conductivity in a system with well-separated spin branches has been calculated in Ref.~\onlinecite{Large_R_classical}, we re-derive it in this Appendix, in order to introduce the kinetic equation with quantum corrections that lead to the interference contribution to the conductivity, see Appendix \ref{AppendixB} below.

The distribution function of electrons with energy $E=E_{\rm F}$ in the two-band model can be written as
\begin{equation}
	f= \sum_{s=\pm} A_s(\varphi)\delta[ E_{\rm F}- E_s(k)],
\end{equation}
so that the current reads
\begin{equation}
	j_x=ev_{\rm F} \sum_{s=\pm}  g_s \big \langle A_s(\varphi) \cos\varphi   \big \rangle_\varphi.
\end{equation}
Here $v_{\rm F}$ is the Fermi velocity, Eq.~(\ref{vF}), equal in both branches, the angular brackets denote averaging over $\varphi$, and the densities of states at the Fermi energy in the branches are given by Eq.~\eqref{g}.

Within the Drude-Boltzmann  approximation,  functions $A_s(\varphi)$ obey the system of two coupled  balance equations:
\begin{align}
\label{kin}
&	-e{\cal E}v_{\rm F}\cos\varphi \\
& = \sum_{s'=\pm}  \big \langle \Gamma_{ss'}^{\rm in} (\varphi-\varphi')A_{s'}(\varphi') -\Gamma_{s's}^{\rm out}(\varphi-\varphi') A_s(\varphi)   \big \rangle_{\varphi'} . \nonumber
\end{align}
Here ${\cal E}$ is the electric field, and we have introduced the ingoing and outgoing scattering rates,
\begin{equation}
	 \Gamma_{ss'}^{\rm in} (\theta)=\frac{2\pi}{\hbar}\left|  V_{ss'}(\theta)\right|^2 g_{s'},
	\quad
	\Gamma_{ss'}^{\rm out} (\theta)=\frac{2\pi}{\hbar}\left|  V_{ss'}(\theta)\right|^2 g_{s},
\end{equation}
which are related as follows: $ \Gamma_{ss'}^{\rm in} =\Gamma_{s's}^{\rm out}$.
The total outgoing rates, Eq.~\eqref{tau}, coincide for two bands
\begin{equation}
	\frac{1}{\tau} = \big\langle  \Gamma_{++}^{\rm out}+ \Gamma_{-+}^{\rm out}\big \rangle_\theta=\big\langle  \Gamma_{--}^{\rm out}+ \Gamma_{+-}^{\rm out}\big \rangle_\theta= \frac{2\pi}{\hbar} \frac{g_++g_-}{2} |V_0|^2.
\end{equation}
 We search for solution of Eq.~\eqref{kin} in the form
 \begin{equation}
 \label{As}
	 A_s= a_s e{\cal E}l\cos\varphi,
 \end{equation}
 where
$a_s$ are  dimensionless coefficients.   In terms of  $a_s$ the Drude conductivity reads
 \begin{equation}
\label{sigmaVK}
	   \sigma_{\rm D}=\frac{e^2v_{\rm F}^2 \tau}{2}\sum_{s = \pm} g_s a_s.
 \end{equation}
Substituting Eq.~\eqref{As} into Eq.~\eqref{kin} we see that $a_s$ obey the following set of coupled equations:
\begin{equation}
	  1=a_s- \sum_{s'}\gamma_{ss'}a_{s'},
 \label{system-as0}
\end{equation}
where
\begin{equation}
  \gamma_{ss'}= \tau \big \langle \Gamma_{ss'}^{\rm in} (\theta) \cos\theta \big \rangle_\theta.
\end{equation}
  Simple calculation yields:
	\begin{align}
  \label{gamma}
		  \gamma_{++}=-\gamma_{-+}= \frac{g_+}{2(g_++g_-)}, \\
			\gamma_{--}=- \gamma_{+-}=\frac{g_-}{2(g_++g_-)}, \nonumber
	\end{align}
and 
\begin{equation}
\label{a-s}
	a_s=\frac{2 g_s}{g_++g_-}.
\end{equation}
Finally, the Drude conductivity becomes
\begin{equation}
	  \sigma_{\rm D}= e^2v_{\rm F}^2\tau \frac{g_+^2+g_-^2}{g_++g_-}=e^2v_{\rm F}^2\tau \frac{m}{2\pi\hbar^2}(1+R^2).
\end{equation}

We note that for the fixed value of $E_{\rm F}$, the Fermi velocity depends on $R$ as
\begin{equation}
v_{\rm F}^2=\frac{2 E_{\rm F}}{m}\frac{1}{1-R^2}.
\end{equation}
As a result, we get
\begin{equation}
\left.\sigma_{\rm D}\right|_{E_{\rm F}=\text{const}}=\frac{e^2}{\pi} E_{\rm F}\tau \frac{1+R^2}{1-R^2}.
\label{sigmaD-R}
\end{equation}

In fact, experimentally, it is the electron concentration that is fixed.
When the Fermi level is located above the Dirac point, the electron concentration
is given by
\begin{equation}
	n = \frac{\left( k_{\rm F}^{+}\right)^2 
	+ \left( k_{\rm F}^{-}\right)^2}{4\pi} = \frac{m}{ \pi\hbar^2} 
	\left( E_{\rm F} + \frac{m\alpha^2}{\hbar^2}\right),
\end{equation}
where the Fermi wavevectors in the subbands are given by Eq.~\eqref{kF}.

From Eq.~\eqref{vF}
we express the Fermi velocity in terms of the fixed concentration and the parameter $R$:
\begin{equation}
v_{\rm F}^2 =\frac{2\pi \hbar^2 n}{m^2(1+R^2)} .
\end{equation}
As a result,
the Drude conductivity for a fixed value of $n$ takes its conventional form:
\begin{equation}
\sigma_{\rm D}=
\frac{e^2 n \tau}{m}.
\label{Drude-n}
\end{equation}
Thus, the Drude conductivity for fixed electron concentration does not depend on $R$.

Writing the diffusion equations for 2D concentrations in two subbands, $n_\pm$, and noting that they are related as $n_+/n_- = g_+/g_-$, we obtain that
the diffusion coefficient is also $R$-independent:
\begin{equation}
\text{D}=\frac{v_{\rm F}^2\tau}{2}(1+R^2) = \frac{\pi \hbar^2 n \tau}{m^2}.
\label{Dne}
\end{equation}

\section{Quantum corrections to kinetic equation}
\label{AppendixB}

\begin{figure}[t]
\includegraphics[width=0.8\linewidth]{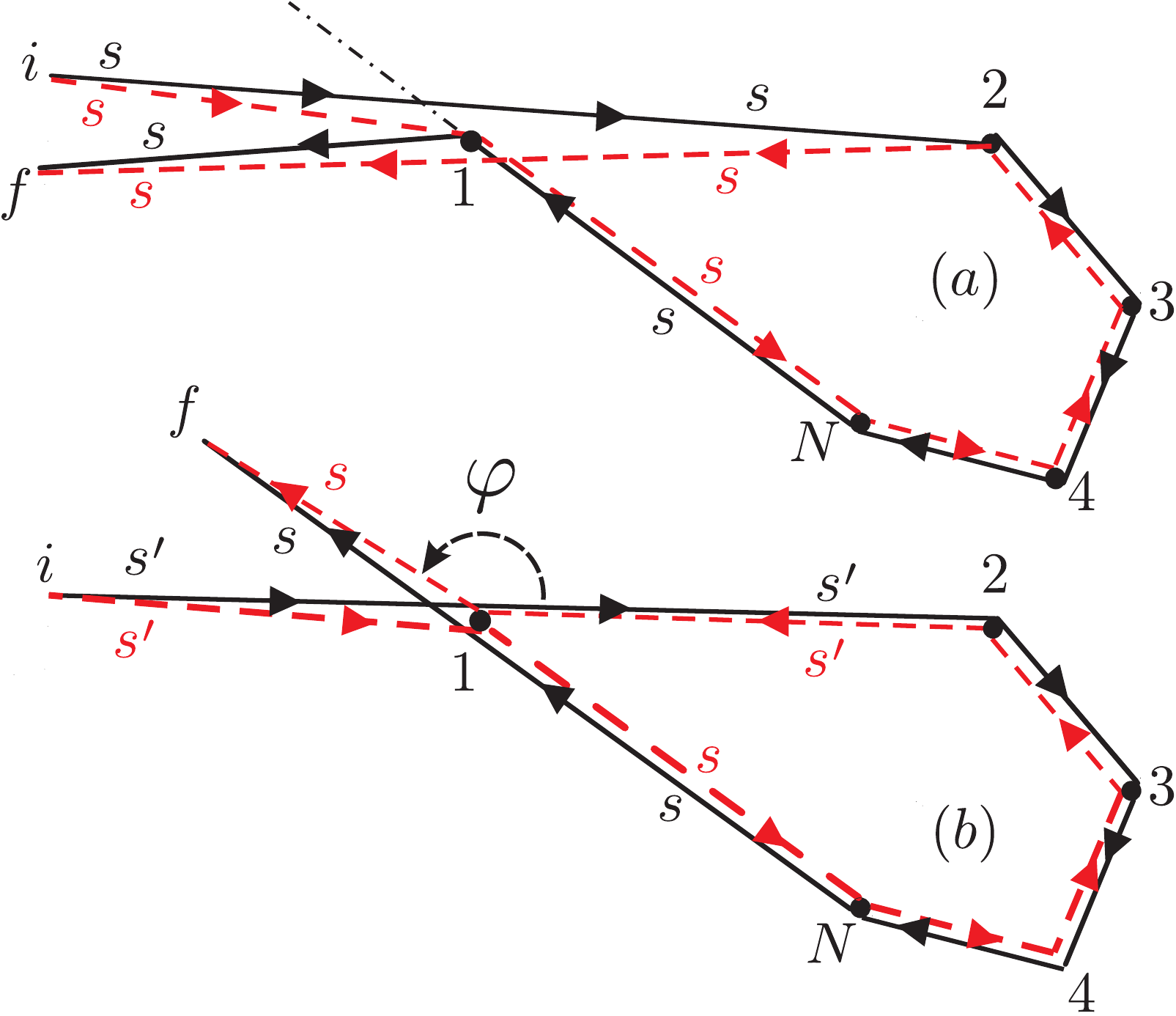}
\caption{Trajectories with $N$ impurities giving rise to the quantum conductivity correction. The processes of backscattering (a) and scattering by an arbitrary angle $\varphi$ (b) contribute to $\sigma_\text{bs}$ and $\sigma_\text{non-bs}$, respectively.}
\label{fig:traject_diagr}
\end{figure}

   As it was shown  in Ref.~\onlinecite{DKG97}, the weak localization correction to the conductivity can be interpreted in terms of localization-induced correction to scattering cross section on a single impurity.   Below we generalize this approach for the system with the strong Rashba splitting of the spectrum. The corresponding trajectories are presented in Fig.~\ref{fig:traject_diagr}.
   
  Within the  kinetic equation formalism, weak localization leads to corrections  to the ingoing scattering rates, so that Eq.~\eqref{system-as0} modifies
   \be
  1=a_s- \sum_{s'}(\gamma_{ss'}+\delta \gamma_{ss'}) a_{s'}, ~~\,\,\, s=\pm.
 \label{system-as2}
  \ee
  Here $\gamma_{ss'}$ are given by Eq.~\eqref{gamma},  
	\be \delta \gamma_{ss'}= \tau {\big \langle \delta \Gamma_{ss'}^{\rm in} (\varphi) \cos\varphi \big \rangle_\varphi},\ee
  and $\delta \Gamma_{ss'}^{\rm in}$ are the interference-induced corrections to the scattering rates.  Treating the  correction $\delta \gamma_{ss'} a_{s'}$ as a small perturbation, one can replace in this term  coefficients $a_s$ with their Drude values given by Eq.~\eqref{a-s}. Doing so, we find that  the weak localization induced corrections to $a_s$ obey
      \be
  \left \{\begin{array}{c}
            \displaystyle \lambda_+= \delta a_+ +\frac{g_- \delta a_- - g_+\delta a_+ }{2(g_++g_-)}, \vspace{2mm} \\
          \displaystyle \lambda_-= \delta a_- +\frac{g_+ \delta a_+ - g_-\delta a_- }{2(g_++g_-)},
          \end{array}
      \right.
  \label{system-as1}
  \ee
  where
  \be
  \lambda_{s}= \frac{2}{g_++g_-}\sum_{s'} \delta \gamma_{ss'}g_{s'}.
  \ee
The conductivity correction is proportional to $\sum_s g_s\delta a_s.$  Solving  Eq.~\eqref{system-as1}, we get
  \be
  \sum_s g_s\delta a_s=  \frac{2}{g_++g_-}\sum_{s}  \lambda_s g_s^2.
  \label{das}
  \ee
  Substituting Eq.~\eqref{das} into Eq.~\eqref{sigmaVK} we express the conductivity correction via corrections to the scattering rates
  \be
  \sigma(B=0)=\frac{2e^2 l^2}{(g_++g_-)^2}  \sum_{ss'} g_s^2g_{s'} \big \langle
  \delta \Gamma_{ss'}^{\rm in} (\varphi) \cos\varphi \big \rangle.
  \label{sigmaWAL}
  \ee
  Similar to Refs.~\onlinecite{DKG97}  and~\onlinecite{GKO}  we express $\delta \Gamma_{ss'}^{\rm in} (\varphi)$ in terms of return probability $w_{ss'}(\varphi), $ which depends now on  the branch indices ($s'$ initial branch, $s$ final branch). To this end, we  introduce rates
  \be
  \nu_{ss'}(\varphi)=l^2w_{ss'} (\varphi) g_{s'} W_{s's}(\pi-\varphi)
 \label{nu}
  \ee
  (rule of summation over repeated indices does not apply here).
  Correction $\delta \Gamma_{ss'}^{\rm in} (\varphi)$ is given by the sum of so-called backscattering and non-backscattering corrections~\cite{DKG97,GKO}:
 \be
 \delta \Gamma_{ss'}^{\rm in}(\varphi)\!=\!C_0\!\left[ 2\pi\delta_{ss'}\delta(\varphi-\pi)\sum_{s''}\!\big \langle \nu_{ss''}(\varphi') \big \rangle_{\varphi'}\! -\nu_{ss'}(\varphi)\right]
\label{dG}
 \ee
with
 \be
 C_0 = \frac{8\pi^2 \hbar }{m v_F l}.
 \label{C-0}
 \ee
 This coefficient  is responsible for the smallness of the quantum correction.

\begin{widetext}
  Substituting Eq.~\eqref{dG} into Eq.~\eqref{sigmaWAL} 
and using
$w_{ss'}=w_{s's}$ and $W_{ss'}=W_{s's}$, we finally get
\begin{align}
&\sigma(B=0)=-\frac{e^2l^2 C_0}{4g_0^2} 
\big\langle \sum_{ss'} g_s g_{s'}(g_s^2\!+\!g_{s'}^2\!+\!2 g_s g_{s'}\cos\varphi) w_{ss'}(\varphi)W_{s's}(\pi-\varphi)\big\rangle_\varphi . 
\end{align}
This expression can be shown to coincide with a sum of Eqs.~\eqref{sigma_a_zero_field} and~\eqref{sigma_b_zero_field} of the main text (see also Supplemental Material \cite{SM}).

\section{Dephasing due to Coulomb interaction}
\label{dephasing}

The equation for the Cooperon in the presence of inelastic 
electron-electron scattering due to the Coulomb repulsion~\cite{AAreview}
reads (see Supplemental Material \cite{SM}):

\begin{equation}
\mathbf{C}_{ss'}(\varphi,\varphi';{\bm q})
=\mathbf{C}_{ss'}^0(\varphi,\varphi';{\bm q})
+\sum_{s_1s_2}\int\frac{d\varphi_1}{2\pi}\int\frac{d\varphi_2}{2\pi}
\mathbf{C}_{ss_1}^0(\varphi,\varphi_1;{\bm q})
\Sigma^\phi_{s_1s_2}(\varphi_1,\varphi_2,{\bm q})
\mathbf{C}_{s_2s'}(\varphi_2,\varphi';{\bm q}).
\label{AppC-Bethe-Salpeter}
\end{equation}
The Cooperon self-energy is, in general, a matrix in subband space,
given by
\begin{eqnarray}
\Sigma^\phi_{s_1s_2}(\varphi_1,\varphi_2)
&=&
16 \pi^2 \tau^4 g_{s_1}g_{s_2}
\int_{-T}^T\frac{d\Omega}{(2\pi)}\int\frac{d^2 Q}{(2\pi)^2}
\frac{T}{\Omega}\text{Im}U(\Omega,{\bm Q})
\mathbf{C}_{s_1s_2}^0(\varphi_1,\varphi_2;\Omega,{\bm Q})
P(\varphi_1;\Omega,\bm Q)P(\varphi_2;\Omega,\bm Q)
.
\label{AppC-self-energy-phi-1}
\end{eqnarray}
\end{widetext}
Here $$U(\Omega,{\bm Q})=\frac{1}{2g_0}\frac{\text{D}Q^2-{\rm i}\Omega}{\text{D}Q^2}$$ is the Fourier component of the dynamically screened Coulomb potential \cite{SM}, $\text{D}$ is the diffusion coefficient, Eq.~\eqref{Dne}, and $g_0=m/(2\pi)$.

The integral over $\Omega$ diverges logarithmically at $\Omega\to 0$.
We regularize this divergence self-consistently in a usual way~\cite{AAreview} at $\Omega$ of the order of the dephasing rate $1/\tau_\phi$ and get
\begin{equation}
{\Sigma}^\phi_{s_1s_2}\simeq
- e^{{\rm i}(\varphi_2-\varphi_1)} \frac{s_1s_2g_{s_1}g_{s_2}}{ g_0 }\ 
\frac{\pi\tau^2}{\tau_\phi}
\label{AppC-SE-final}
\end{equation}
with
\begin{equation}
\frac{1}{\tau_\phi}=\frac{T}{4\pi g_0 \text{D}} \ln(T\tau_\phi)
\simeq \frac{T}{4\pi g_0 \text{D}} \ln(4\pi g_0 \text{D}).
\label{AppC-tau-phi0}
\end{equation}

In the diffusion approximation $ql\ll 1$, the solution of Eq~(\ref{AppC-Bethe-Salpeter}) with the self-energy (\ref{AppC-SE-final}) takes the form~\cite{SM}:
\begin{widetext}
\begin{eqnarray}
\mathbf{C}_{ss'}(\varphi,\varphi';{\bm q}) \simeq
\frac{e^{{\rm i}(\varphi'-\varphi)}}{4\pi g_0\tau^2}
\left\{\frac{ss'+{\rm i}qlR(s'\cos\varphi+s\cos\varphi')}{\text{D}q^2 + \frac{1}{\tau_\phi}}+
2\tau\left[\sin\varphi\sin\varphi'
+
\frac{\frac{\text{D}}{1+R^2}q^2 + \frac{1}{\tau_\phi}}{\text{D}q^2 + \frac{1}{\tau_\phi}}
\cos\varphi\cos\varphi'\right]
\right\}.
\label{AppC-full-Coop-diff}
\end{eqnarray}
\end{widetext}
The logarithmically divergent diffusive term in the conductivity correction is thus cut off by $1/\tau_\phi$ given by Eq.~(\ref{AppC-tau-phi0}).

In the experimentally relevant case, when the electron concentration $n$
is kept fixed by applying the gate voltage, the diffusion coefficient
$\text{D}$ (and hence the ratio $\tau_\phi/\tau$)
do not depend on $R$, see Eqs.~(\ref{Dne}) and (\ref{AppC-tau-phi0}):
\begin{equation}
\frac{\tau}{\tau_\phi} =\frac{m T}{2 \pi n } \ln\frac{2\pi n \tau}{m}.
\end{equation}
As a result, the $T$-dependent  diffusive term Eq.~\eqref{diff_zero_field}
\begin{equation}
\sigma_\text{diff}(T)=\frac{e^2}{4\pi^2 \hbar}\ln\frac{\tau_\phi}{\tau}
\simeq \frac{e^2}{4\pi^2 \hbar}\ln\frac{2\pi\hbar^2 n}{m T \ln(2\pi\hbar n \tau/m)}
\label{AppC-sigma-diff-ne}
\end{equation}
is $R$-independent for fixed $n$.

\newpage

\begin{widetext}

\renewcommand{\cite}[1]{{[}\onlinecite{#1}{]}}

\renewcommand{\thepage}{S\arabic{page}}
\renewcommand{\theequation}{S.\arabic{equation}}
\renewcommand{\thefigure}{S\arabic{figure}}
\renewcommand{\bibnumfmt}[1]{[S#1]}
\setcounter{page}{1}
\setcounter{section}{0}
\setcounter{equation}{0}
\setcounter{figure}{0}

\begin{center}
{\bf {ONLINE SUPPLEMENTAL MATERIAL\\
Weak antilocalization in two-dimensional systems with large Rashba splitting}
}
\end{center}

%
%

\begin{center}
\parbox{16cm}
{\small In Supplemental Material, we present the calculation of the interference-induced correction to the conductivity at zero magnetic field using the chiral-state basis 
as well as the details of the evaluation of the dephasing rate at strong Rashba splitting of the spectrum.}
\end{center}



\begin{center}
{\bf S1. Zero-field correction: Calculation in subband basis and momentum space}
\end{center}
\label{IG_notes}

In this section we present the derivation of the conductivity correction 
in zero magnetic field using the momentum representation of
Green's functions in the basis of chiral subbands. This calculation complements the one of the main text that employed the coordinate representation in the singlet-triplet basis, bridging it with the analysis of the quantum corrections to the kinetic equation (Appendix B). This allows us to clarify the relation between the two alternative approaches to the analysis of quantum interference effects at strong spin-orbit splitting of the spectrum.

\subsection*{1. Cooperon}

We start with deriving the explicit form of the main building block of the conductivity correction -- the Cooperon propagator.
The Cooperon ${\cal C}_{ss'}({\bm k},{\bm k'},{\bm q})$ is a matrix in the basis of subbands $s=\pm$ s defined in Fig.~\ref{fig:coop}.
\begin{figure}[ht]
\centerline{\includegraphics[width=14cm]{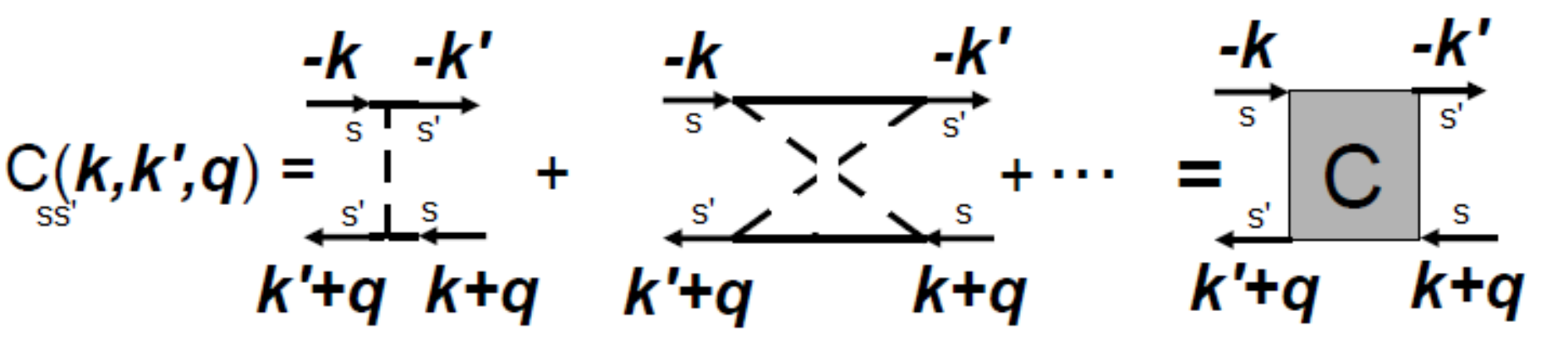}}
\caption{Diagram for Cooperon. The outer lines are not included into the Cooperon and only show the directions of arrows here.}
\label{fig:coop}
\end{figure}
The arguments of the Cooperon are read off from the arguments of the advanced Green functions (lower line) as shown in Fig.~\ref{fig:coop}:
the first momentum $\bm k$ and first index $s$ correspond to the incoming arrow, the second momentum $\bm k'$ and second index $s'$ correspond
to the outgoing arrow for the lower line in the diagram. We note that here
the Cooperon ladder starts with a single impurity line, unlike in the main
text, where the minimum Cooperon contains two impurity lines, see Fig. 2 and
Eq.~(11) of the main text.

Introducing the angles
\[
\varphi=\varphi_{\bm k}-\varphi_{\bm q}, \qquad
\varphi'=\varphi_{\bm k'}-\varphi_{\bm q}
\]
we obtain the following equations for Cooperons:
\begin{align}
\label{C_system1}
	{\cal C}_{++}(\varphi,\varphi') = W_+(\varphi-\varphi')  + \frac{2\pi\tau}{\hbar}\left<W_+(\varphi-\varphi_1)  g_+  P(\varphi_1) {\cal C}_{++}(\varphi_1,\varphi')
	+ W_-(\varphi-\varphi_1)  g_-  P(\varphi_1) {\cal C}_{-+}(\varphi_1,\varphi')\right>_{\varphi_1}, \\
\label{C_system2}
	{\cal C}_{-+}(\varphi,\varphi') = W_-(\varphi-\varphi')  + \frac{2\pi\tau}{\hbar}\left<W_-(\varphi-\varphi_1)  g_+  P(\varphi_1) {\cal C}_{++}(\varphi_1,\varphi')
	+ W_+(\varphi-\varphi_1)  g_-  P(\varphi_1) {\cal C}_{-+}(\varphi_1,\varphi')\right>_{\varphi_1}.
\end{align}
Here
\begin{equation}
	P(\varphi) =  \frac{1}{1 - {\rm i}ql\cos{\varphi} + \tau/\tau_\phi}, \qquad l = v_{\rm F}\tau,
\label{Pphi}
\end{equation}
and
\begin{align}
\label{corr}
W_{+}(\theta) = |V_{++}|^2 = V_0^2\left(\frac{1+{\rm e}^{{\rm i}\theta}}{2}\right)^2 = V_0^2 \frac{1+2{\rm e}^{{\rm i}\theta} + {\rm e}^{2{\rm i}\theta}}{4}
=\frac{V_0^2}{2}{\rm e}^{{\rm i}\theta} (1+\cos\theta), \\
W_{-}(\theta) = |V_{+-}|^2 = V_0^2\left(\frac{1-{\rm e}^{{\rm i}\theta}}{2}\right)^2=V_0^2 \frac{1-2{\rm e}^{{\rm i}\theta} + {\rm e}^{2{\rm i}\theta}}{4}
=\frac{V_0^2}{2}{\rm e}^{{\rm i}\theta} (-1+\cos\theta).
\end{align}
Equations for ${\cal C}_{--}$ and ${\cal C}_{+-}$ are obtained by exchanging $g_+ \leftrightarrow g_-$.

It is convenient to single-out the phase factor in the Cooperons:
\begin{equation}
{\cal C}_{ss'}(\varphi,\varphi')={\rm e}^{{\rm i}(\varphi-\varphi')} \tilde{C}_{ss'}(\varphi,\varphi').
\label{tildeC}
\end{equation}
Then we rewrite the Cooperon equations in the following form:
\begin{eqnarray}
	\tilde{C}_{++}(\varphi,\varphi') &=& \frac{V_0^2}{2}[1+\cos(\varphi-\varphi')] \nonumber
\\
&+&
\left\langle  \frac{\tau}{2\tau_+}[1+\cos(\varphi-\varphi_1)] P(\varphi_1) \tilde{C}_{++}(\varphi_1,\varphi')
	+ \frac{\tau}{2\tau_-}[-1+\cos(\varphi-\varphi_1)]  P(\varphi_1) \tilde{C}_{-+}(\varphi_1,\varphi')\right\rangle_{\varphi_1},
\label{Cpp}
\\
	\tilde{C}_{-+}(\varphi,\varphi') &=& \frac{V_0^2}{2}[-1+\cos(\varphi-\varphi')] \nonumber
\\
&+&
\left\langle  \frac{\tau}{2\tau_+}[-1+\cos(\varphi-\varphi_1)] P(\varphi_1) \tilde{C}_{++}(\varphi_1,\varphi')
	+ \frac{\tau}{2\tau_-}[1+\cos(\varphi-\varphi_1)]  P(\varphi_1) \tilde{C}_{-+}(\varphi_1,\varphi')\right\rangle_{\varphi_1}.
	\label{Cmp}
\end{eqnarray}
Here we have introduced 
\begin{equation}
	\frac{1}{\tau_\pm} = \frac{2\pi}{\hbar}V_0^2 g_\pm = \frac{1\mp R}{\tau}.
\end{equation}
The explicit solution of Eqs.~(\ref{Cpp}) and (\ref{Cmp}) has the form:
\begin{eqnarray}
\frac{2}{V_0^2}\tilde{C}_{++}(\varphi,\varphi')&=&
\frac{1}{d_2}\left( 1 - \frac{P_0 + P_2}{2}\right) -
   \frac{P_1 R}{d_2} (\cos\varphi + \cos\varphi') + \frac{1 - P_0}{d_2} \cos\varphi\cos\varphi'
    + \frac{1}{d_1}\sin\varphi\sin\varphi',
    \\
\frac{2}{V_0^2}\tilde{C}_{-+}(\varphi,\varphi')&=&
-\frac{1}{d_2}\left( 1 - \frac{P_0 + P_2}{2}\right) -
   \frac{P_1 R}{d_2} (\cos\varphi - \cos\varphi') + \frac{1 - P_0}{d_2} \cos\varphi\cos\varphi'
    + \frac{1}{d_1}\sin\varphi\sin\varphi',
    \label{tildeCss'}
\end{eqnarray}
Here
\begin{equation}
\label{Pn}
P_n = \left< P(\varphi) \cos{n\varphi}\right>_\varphi
\end{equation}
and
\begin{equation}
d_1=1-\frac{P_0-P_2}{2}, \qquad d_2=(1 - P_0) \left(1 - \frac{P_0 + P_2}{2}\right) - P_1^2 R^2.
\label{d1d2}
\end{equation}
The solutions for $\tilde{C}_{--}$ and $\tilde{C}_{+-}$ are obtained by the replacement $R\to -R$, which only changes the sign
of the $P_1 R$ term.

The full Cooperon ${\cal C}_{ss'}$ can be conveniently represented in the matrix form
\begin{equation}
{\cal C}_{ss'}(\varphi,\varphi')=\frac{V_0^2}{2}{\rm e}^{{\rm i}(\varphi-\varphi')}
\left( \begin{array}{c}
	s\\
\cos\varphi  \\
\sin\varphi
	\end{array} \right)^T
	\left(\hat{\cal I}-\hat{\cal B}\right)^{-1}
	\left( \begin{array}{c}
	s'\\
\cos\varphi'  \\
\sin\varphi'
	\end{array} \right),
\label{Css'}
\end{equation}
where $\hat{\cal I}$ is a $3\times 3$ unit matrix and
\begin{equation}
	\hat{\cal B} = \left( \begin{array}{ccc}
	P_0 & - RP_1 & 0 \\
	- RP_1 & (P_0+P_2)/2 & 0\\
	0&  0 &  (P_0-P_2)/2
	\end{array} \right)
\label{MatrixB}
\end{equation}
is a block-diagonal matrix.
Similarly, we can write the disorder correlator in the same basis as
\begin{equation}
 W_{ss'}(\varphi,\varphi')=\frac{V_0^2}{2}{\rm e}^{{\rm i}(\varphi-\varphi')}
\left( \begin{array}{c}
	s\\
\cos\varphi  \\
\sin\varphi
	\end{array} \right)^T
	\hat{\cal I}
	\left( \begin{array}{c}
	s'\\
\cos\varphi'  \\
\sin\varphi'
	\end{array} \right).
\label{Wss'}
\end{equation}

The result (\ref{Css'}) can be equivalently cast 
in terms of exponential angular harmonics as follows:
\begin{equation}
	 {\cal C}_{ss'}(\varphi,\varphi') =
	\frac{V_0^{2}}{ 2}\left( \begin{array}{c}
	1\\ s {\rm e}^{{\rm i}\varphi}  \\ {\rm e}^{2{\rm i}\varphi}
	\end{array} \right)^T
	\left(\hat{\cal I}-\hat{A}\right)^{-1} \hat{m}
	\left( \begin{array}{c}
	1\\ s' {\rm e}^{-{\rm i}\varphi'}  \\ {\rm e}^{-2{\rm i}\varphi'}
	\end{array} \right)
\end{equation}
with
\begin{equation}
\hat{A} = \left( \begin{array}{ccc}
	P_0/ 2 & - RP_1/2 & P_2/2 \\
	- RP_1 & P_0 & -R P_1\\
	P_2/2&  -R P_1/2 &  P_0/2
	\end{array} \right), \qquad
\hat{m} = \left( \begin{array}{ccc}
	1/2 & 0 & 0 \\
	0 & 1 & 0\\
	0&  0 &  1/2
	\end{array} \right).
\end{equation}

For the calculation of the backscattering and non-backscattering contributions to the conductivity
we will need the Cooperons $ {\cal C}_{ss'}^{(3)}(\varphi,\varphi') $ and $ {\cal C}_{ss'}^{(2)}(\varphi,\varphi') $ starting with 3 and 2 impurity lines, respectively. These are given by
\begin{align}
{\cal C}_{ss'}^{(p)}(\varphi,\varphi')=\frac{V_0^2}{2}{\rm e}^{{\rm i}(\varphi-\varphi')}
\left( \begin{array}{c}
	s\\
\cos\varphi  \\
\sin\varphi
	\end{array} \right)^T
	{\hat{\cal B}}^{p-1}\left(\hat{\cal I}-\hat{\cal B}\right)^{-1}
	\left( \begin{array}{c}
	s'\\
\cos\varphi'  \\
\sin\varphi'
	\end{array} \right)
\label{Css'p} \\ 
	=
	\frac{V_0^{2}}{ 2}\left( \begin{array}{c}
	1\\ s {\rm e}^{{\rm i}\varphi}  \\ {\rm e}^{2{\rm i}\varphi}
	\end{array} \right)^T
	\hat{A}^{p-1}\left(\hat{\cal I}-\hat{A}\right)^{-1} \hat{m}
	\left( \begin{array}{c}
	1\\ s' {\rm e}^{-{\rm i}\varphi'}  \\ {\rm e}^{-2{\rm i}\varphi'}
	\end{array} \right). \nonumber
\end{align}

\subsection*{2. Backscattering contribution to the conductivity}

The backscattering conductivity correction is described by the diagram in Fig.~\ref{fig:back}.
The backscattering correction contains only the diagonal Cooperons and is given by (hereafter we put $\hbar=1$ and restore $\hbar$ in the final expressions)
\begin{equation}
	\sigma_\text{bs} = \frac{e^2}{4\pi} \sum_{s=\pm} \int\frac{d^2q}{(2\pi)^2} \int\frac{d^2k}{(2\pi)^2} {\bm v_s}({\bm k}){\bm v_s}(-{\bm k}+{\bm q})
G^R_{s}(\bm k)G^A_{s}(\bm k)\,  G^R_{s}(-\bm k + \bm q)G^A_{s}(-\bm k+\bm q) \, {\cal C}_{ss}^{(3)}(-\bm k,\bm k,\bm q).
\label{bs1}
\end{equation}
Here the renormalized velocity is given by
\begin{equation}
	\bm v_s(\bm k) = v_{\rm F} \frac{\bm k }{ k} \frac{\tau_{\rm tr}^{(s)} }{ \tau}.
\label{vs-tr}
\end{equation}
\begin{figure}[ht]
\centerline{\includegraphics[width=8cm]{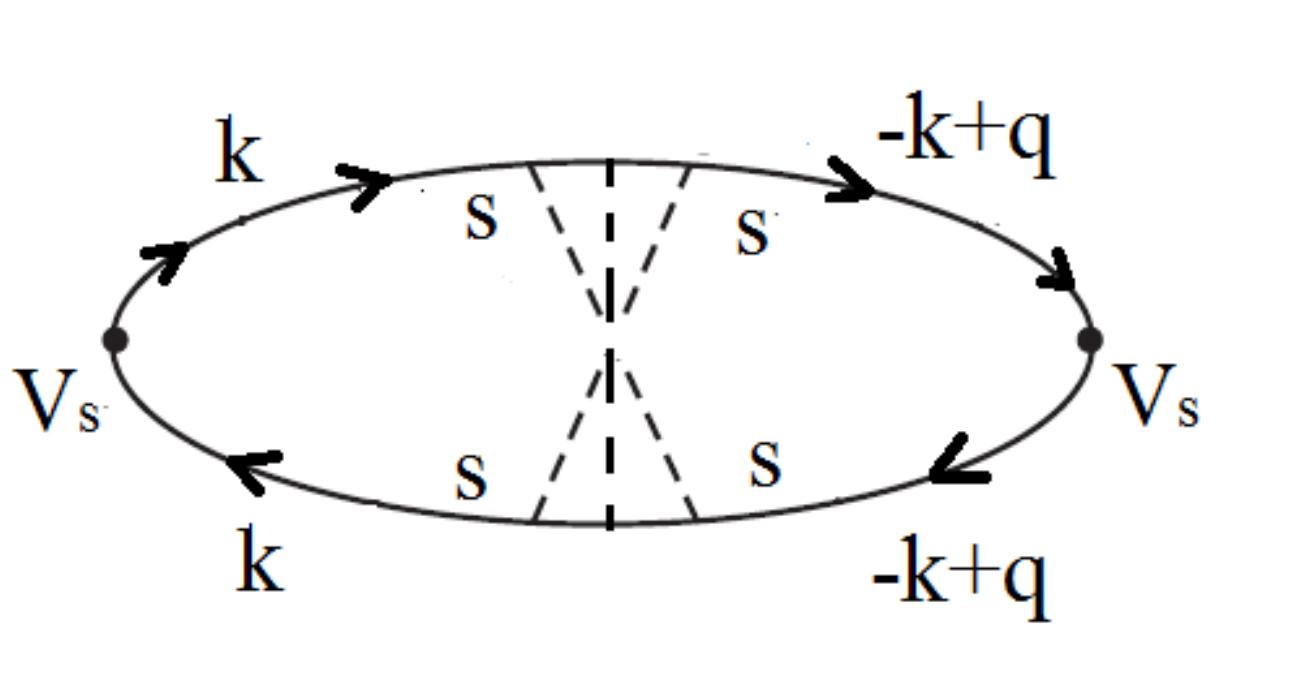}}
\caption{Diagram for the backscattering contribution to the conductivity correction. The lowest-order (three-impurity) term is shown.}
\label{fig:back}
\end{figure}
Using the identity
\begin{equation}
G^RG^A={\rm i}\tau (G^R-G^A),
\label{identity}
\end{equation}
neglecting $\bm q$ in the current vertex,
writing ${\cal C}_{ss}^{(3)}(-\bm k,\bm k,\bm q)={\cal C}_{ss}^{(3)}(\varphi+\pi,\varphi)$,
and retaining only the retarded-advanced product of Green's functions,
we get
\begin{eqnarray}
\sigma_\text{bs}&=&- \frac{e^2 v_F^2 \tau^2}{4\pi}  \sum_{s=\pm} \left[\frac{\tau_{\rm tr}^{(s)}}{\tau}\right]^2\!\!
\int\frac{d^2q}{(2\pi)^2}\!\! \int\frac{d^2k}{(2\pi)^2}  \frac{\bm k }{ k}\frac{(-\bm k) }{ k}
\left[G^R_{s}(\bm k)-G^A_{s}(\bm k)\right]\, {\cal C}_{ss}^{(3)}(\varphi+\pi,\varphi)\,  \left[G^R_{s}(-\bm k + \bm q)-G^A_{s}(-\bm k+\bm q)\right]
\nonumber
\\
&=&-\frac{e^2 l^2}{4\pi}  \sum_{s=\pm} \left[\frac{\tau_{\rm tr}^{(s)}}{\tau}\right]^2
\int\frac{d^2q}{(2\pi)^2} \int\frac{d^2k}{(2\pi)^2}
\left[G^R_{s}(\bm k) G^A_{s}(-\bm k+\bm q)+G^A_{s}(\bm k) G^R_{s}(-\bm k+\bm q)\right] {\cal C}_{ss}^{(3)}(\varphi+\pi,\varphi)
.
\label{bs2}
\end{eqnarray}

In Eq.~(\ref{bs2}), we have the following products of retarded and advanced Green's functions:
$$G^R_{s}(\bm k) G^A_{s}(-\bm k + \bm q), \qquad G^R_{s}(-\bm k+\bm q)G^A_{s}(\bm k).$$
Since the Cooperon does not depend on the absolute value
of $\bm k$, we integrate the products of Green's functions over $|\bm k|$ for the fixed value
 of $\varphi=\varphi_{\bm k}-\varphi_{\bm q}$, using
 $$E_s(\bm p)\simeq E_F+v_F \left(|\bm p|-k_F^{(s)}\right).$$
  We get:
\begin{equation}
	\int \frac{k dk}{2\pi} G^R_{s}(\bm k)G^A_{s}(-\bm k+\bm q)=\frac{1}{V_0^2}\frac{\tau}{\tau_s} P(\varphi+\pi),
\label{PGRGA}
\end{equation}
and, analogously,
\begin{equation}
	\int \frac{kdk}{2\pi} G^R_{s}(-\bm k+\bm q) G^A_{s}(\bm k)=\frac{1}{V_0^2} \frac{\tau}{\tau_{s}} P(\varphi).
\label{PGAGR}
\end{equation}
Substituting Eqs.~(\ref{PGRGA}) and (\ref{PGAGR}) into Eq.~(\ref{bs2}), we obtain
\begin{equation}
\sigma_\text{bs}=-\frac{e^2 l^2}{4\pi}  \sum_{s=\pm} \left[\frac{\tau_{\rm tr}^{(s)}}{\tau}\right]^2\frac{\tau}{\tau_{s}}
\int\frac{d^2q}{(2\pi)^2} \frac{1}{V_0^2} \left\langle\left[P(\varphi+\pi)+P(\varphi)\right]
{\cal C}_{ss}^{(3)}(\varphi+\pi,\varphi)\right\rangle_\varphi.
\label{bs3}
\end{equation}
Using the explicit form of ${\cal C}_{ss}^{(3)}(\varphi+\pi,\varphi)$ from Eq.~(\ref{Css'p}),
one can 
see that it does not depend on $s$.
Thus, the summation over $s$ involves only
$$\left[\frac{\tau_{\rm tr}^{(s)}}{\tau}\right]^2\frac{\tau}{\tau_{s}}=(1-s R)^3,$$
which yields
\begin{equation}
\sum_{s=\pm}\left[\frac{\tau_{\rm tr}^{(s)}}{\tau}\right]^2\frac{\tau}{\tau_{s}}=(1-R)^3+(1+R)^3=2(1+3R^2).
\end{equation}
We also note that the Cooperon ${\cal C}_{++}^{(3)}(\varphi+\pi,\varphi)$ depends only on the absolute value of $q$.
Therefore, one can replace $P(\varphi+\pi)+P(\varphi)$ in Eq.~(\ref{bs3}) by $2P(\varphi)$.
Indeed, $P(\varphi+\pi;{\bm q})=P(\varphi;-{\bm q})$.

At this point it is worth making the connection to Ref.~\onlinecite{GKO},
where the conductivity correction for massive Dirac fermions was calculated in the chiral basis.
The Cooperon ${\cal C}_{++}^{(3)}(\varphi',\varphi)$ that starts with three impurity lines
can be rewritten by using Eq.~(\ref{C_system1}) in two equivalent forms involving ${\cal C}_{ss'}^{(2)}$
\begin{eqnarray}
{\cal C}_{++}^{(3)}(\varphi',\varphi)&=&2\pi\tau\int_0^{2\pi}\frac{d\varphi_1}{2\pi}\left[W_{++}(\varphi',\varphi_1)g_+ P(\varphi_1)
{\cal C}_{++}^{(2)}(\varphi_1,\varphi)
+W_{+-}(\varphi',\varphi_1)g_- P(\varphi_1){\cal C}_{-+}^{(2)}(\varphi_1,\varphi)\right]
\\
&=&
2\pi\tau\int_0^{2\pi}\frac{d\varphi_1}{2\pi}\left[{\cal C}_{++}^{(2)}(\varphi',\varphi_1)g_+ P(\varphi_1) W_{++}(\varphi_1,\varphi)
+ {\cal C}_{+-}^{(2)}(\varphi',\varphi_1)g_- P(\varphi_1)W_{-+}(\varphi_1,\varphi) \right].
\end{eqnarray}
Then the product $\left[P(\varphi')+P(\varphi)\right] {\cal C}_{ss}^{(3)}(\varphi',\varphi)$
in the backscattering correction (\ref{bs3}) (where $\varphi'=\varphi+\pi$) can be cast in the form (here ${\bar s}=-s$)
\begin{eqnarray}
\left[P(\varphi')+P(\varphi)\right] {\cal C}_{ss}^{(3)}(\varphi',\varphi)
&=&2\pi\tau\int_0^{2\pi}\frac{d\varphi_1}{2\pi}\left[P(\varphi'){\cal C}_{ss}^{(2)}(\varphi',\varphi_1)P(\varphi_1) g_sW_{s}(\varphi_1-\varphi)
\right.
\nonumber\\
&+&\left. P(\varphi'){\cal C}_{s{\bar s}}^{(2)}(\varphi',\varphi_1) P(\varphi_1) g_{\bar s}W_{{\bar s}}(\varphi_1-\varphi) \right]
\nonumber
\\
&+&
2\pi\tau\int_0^{2\pi}\frac{d\varphi_1}{2\pi}\left[g_sW_{s}(\varphi'-\varphi_1) P(\varphi_1)
{\cal C}_{ss}^{(2)}(\varphi_1,\varphi)P(\varphi)
\right.
\nonumber\\
&+&\left.
g_{\bar s} W_{{\bar s}}(\varphi'-\varphi_1) P(\varphi_1)
{\cal C}_{{\bar s}s}^{(2)}(\varphi_1,\varphi)P(\varphi)\right].
\end{eqnarray}
Introducing the return probabilities as in Ref.~\onlinecite{GKO},
\begin{equation}
w_{ss'}(\phi-\phi')=\frac{1}{2\pi V_0^2}\int\frac{qdq}{2\pi}\left\langle P(\phi-\phi_q)
{\cal C}_{ss'}^{(2)}(\phi-\phi_q,\phi'-\phi_q;q)P(\phi'-\phi_q)
\right\rangle_{\phi_q},
\label{GKOwss'}
\end{equation}
using $2\pi g_0 \tau=1/V_0^2$ with $g_0=(g_++g_-)/2$,
and performing the integration over $\bm q$ in Eq.~(\ref{bs3}), we can rewrite the backscattering correction in terms
of $w_{ss'}$:
\begin{eqnarray}
\sigma_\text{bs}&=&-\frac{e^2 l^2}{2 V_0^2}
\sum_{s=\pm} (1-s R)^3
 \nonumber
\\
&\times&
\left[
(1-sR)
\left\langle
w_{ss}(\pi+\varphi-\varphi_1)W_{s}(\varphi_1-\varphi)
\right\rangle_{\varphi,\varphi_1}
+
(1+sR)
\left\langle
w_{s\bar{s}}(\pi+\varphi-\varphi_1)W_{\bar{s}}(\varphi_1-\varphi)\right\rangle_{\varphi,\varphi_1}
\right.
\nonumber
\\
&+&
\left.
(1-sR)
\left\langle
W_{s}(\varphi+\pi-\varphi_1) w_{ss}(\varphi_1-\varphi)\right\rangle_{\varphi,\varphi_1}
+
(1+sR)
\left\langle
W_{\bar{s}}(\varphi+\pi-\varphi_1) w_{\bar{s}s}(\varphi_1-\varphi)
\right\rangle_{\varphi,\varphi_1}
\right].
\label{bs3-wW}
\end{eqnarray}
Shifting the angles $\varphi+\pi\to \varphi$ in the first two terms in the square brackets,
we introduce $\theta=\varphi_1-\varphi$ and get
\begin{eqnarray}
\sigma_\text{bs}&=&-\frac{e^2 l^2}{2 V_0^2}
\sum_{s=\pm} (1-s R)^3
 \nonumber
\\
&\times&
\left[
(1-sR)
\left\langle
W_{s}(\pi+\theta)w_{ss}(-\theta)
\right\rangle_{\theta}
+
(1+sR)
\left\langle
W_{\bar{s}}(\pi+\theta)w_{s\bar{s}}(-\theta)
\right\rangle_{\theta}
\right.
\nonumber
\\
&+&
\left.
(1-sR)
\left\langle
W_{s}(\pi-\theta) w_{ss}(\theta)\right\rangle_{\theta}
+
(1+sR)
\left\langle
W_{\bar{s}}(\pi-\theta) w_{\bar{s}s}(\theta)
\right\rangle_{\theta}
\right]
\nonumber
\\
&=&-\frac{e^2 l^2}{V_0^2}
\sum_{s=\pm}
\left[
(1-sR)^4
\left\langle
W_{s}(\pi-\theta) w_{ss}(\theta)\right\rangle_{\theta}
+
(1-sR)^2(1-R^2)
\left\langle
W_{\bar{s}}(\pi-\theta) w_{\bar{s}s}(\theta)
\right\rangle_{\theta}
\right].
\label{bs3-w}
\end{eqnarray}
Finally, restoring the factors $\tau_{\rm tr}^{(s)}/\tau$ and $g_s/g_0$, as well as $\hbar$,
we rewrite Eq.~(\ref{bs3-w}) as
\begin{eqnarray}
\sigma_\text{bs}=-\frac{e^2}{\hbar}l^2
\sum_{s=\pm}
\left[\frac{\tau_{\rm tr}^{(s)}}{\tau}\right]^2\frac{g_s}{g_0}
\left[
\frac{g_s}{g_0}
\left\langle
\frac{W_{s}(\pi-\theta)}{V_0^2}
w_{ss}(\theta)\right\rangle_{\theta}
+
\frac{g_{\bar s}}{g_0}
\left\langle
\frac{W_{\bar{s}}(\pi-\theta)}{V_0^2} w_{\bar{s}s}(\theta)
\right\rangle_{\theta}
\right].
\label{bs3-w-final}
\end{eqnarray}
This is a generalization of the backscattering term (the one without the cosinus) in Eq.~(66) of Ref.~\onlinecite{GKO}
onto the multiband system. The ratio $W_{s}(\pi-\theta)/V_0^2$ corresponds to $\gamma_C(\pi-\theta)/\gamma$
in Ref.~\onlinecite{GKO}, while the additional factors $g_s/g_0$ come from the non-equal densities of states of initial and final states in the two subbands.

Calculating the angular integral over $\varphi$ in Eq.~(\ref{bs3})
using [see Eq.~\eqref{Pn}]
\begin{equation}
\left< P(\varphi)\right>_\varphi= P_0, \qquad
\left< P(\varphi) \cos^2\varphi \right>_\varphi=\frac{P_0+P_2}{2},
\qquad
\left< P(\varphi) \sin^2\varphi \right>_\varphi=\frac{P_0-P_2}{2},
\label{P0P1P2}
\end{equation}
we get
\begin{eqnarray}
 \frac{1}{V_0^2}\left\langle\left[P(\varphi+\pi)+P(\varphi)\right]
{\cal C}_{++}^{(3)}(\varphi+\pi,\varphi)\right\rangle_\varphi
&=&\frac{P_0-P_2}{2}\left[P_0+P_2+\frac{2}{2-P_0+P_2}
-\frac{2}{(1-P_0)(2-P_0-P_2)-2P_1^2R^2}\right].
\nonumber
\\
\label{angular-bs}
\end{eqnarray}
The backscattering correction can be written in a compact form as
\begin{equation}
\label{sigma_bs_IG}
	\sigma_\text{bs}=-\frac{e^2 l^2}{2\pi}(1+3R^2) \int \frac{d^2q}{(2\pi)^2}{\rm Tr} \left[ \hat{\Pi}  \hat{A}^3 \left( \hat{\cal I}-\hat{A}\right)^{-1} \right],
\qquad \hat{\Pi}={\rm diag}(1,-1,1).
\end{equation}
Taking into account that 
\begin{equation}
P_n=i^{|n|}P_0 \left[\frac{1-P_0(1+\gamma)}{1+P_0(1+\gamma)} \right]^{|n|/2}, \qquad \gamma=\frac{\tau}{\tau_\phi},
\label{PnP0}
\end{equation}
we see that expression~\eqref{sigma_bs_IG} coincides with Eq.~(32) of the main text.

Calculating the trace in Eq.~\eqref{sigma_bs_IG}
and using
$$1+\gamma+q^2l^2=\frac{1}{P_0^2}, \qquad l^2\, q dq =-\frac{dP_0}{P_0^3},$$
we arrive at
\begin{equation}
\sigma_\text{bs}=-\frac{e^2 l^2}{4\pi^2} (1+3R^2)
\int_0^\infty q dq \mathcal{F}_\text{bs}=-\frac{e^2}{4\pi^2} (1+3R^2)
\int_0^{1/(1+\gamma)} \frac{dP_0}{P_0^3} \mathcal{F}_\text{bs},
\label{bs4}
\end{equation}
where
\begin{equation}
\label{angular-bs1}
	\mathcal{F}_\text{bs}=P_0
\left\{2(1+\gamma)\left[\frac{P_0}{1+P_0(1+\gamma)}\right]^2 +\frac{1}{1+P_0\gamma} -\frac{1}{1+P_0\gamma-P_0^2(2-R^2+2\gamma)+P_0^3(1-R^2)(1+\gamma)}\right\}.
\end{equation}
For finite $\gamma$, the analytical integration over $P_0$ requires the solution of the cubic equation,
see the denominator of the last term in Eq.~(\ref{angular-bs1}).
For $\gamma=0$, this denominator factorizes, but the integral in Eq.~(\ref{bs4}) diverges logarithmically at the upper limit $P_0\to 1$,
as it should for the interference correction in two dimensions.
Let us then set $\gamma=0$ in $\mathcal{F}_\text{bs}$ and single out the divergent term.
We find
\begin{equation}
\left.\mathcal{F}_\text{bs}\right|_{\gamma=0}=P_0+\frac{2P_0^3}{(1+P_0)^2}
- \frac{P_0}{(1-P_0)[1+P_0-P_0^2(1-R^2)]}.
\label{Fbs-gamma0}
\end{equation}
The indefinite integral over $P_0$ in Eq.~(\ref{bs4}) with (\ref{Fbs-gamma0}) is evaluated analytically as:
\begin{eqnarray}
\label{indefinite_int}
\int \frac{dP_0}{P_0^3} \mathcal{F}_\text{bs}
&=&\frac{1}{1+R^2}\ln(1-P_0)-\frac{2}{1+P_0}
\nonumber
\\
&-&\frac{1}{2(1+R^2)}\ln[1+P_0-P_0^2(1-R^2)]
+\frac{1+2R^2-2R^4}{(1+R^2)\sqrt{5-4R^2}}\text{arctanh}\left[\frac{1-2P_0(1-R^2)}{\sqrt{5-4R^2}}\right].
\end{eqnarray}
The first term here is the diffusive contribution that
yields  the logarithmic weak-antilocalization correction. For a finite system of size $L$, taking the limits $P_0=0$ and $1-q_\text{min}^2l^2/2$, 
where $q_\text{min}\sim 1/L$, we get
\begin{equation}
\sigma_\text{bs}^\text{diff}
= \frac{e^2}{4\pi^2}\frac{1+3R^2}{1+R^2} \ln\frac{2}{q_\text{min}^2l^2}.
\label{bs-diff}
\end{equation}
The prefactor here depends on $R$, but as we will see below, the diffusive term in the non-backscattering
correction will restore the universality.

\begin{figure}[ht]
\centerline{\includegraphics[width=10cm]{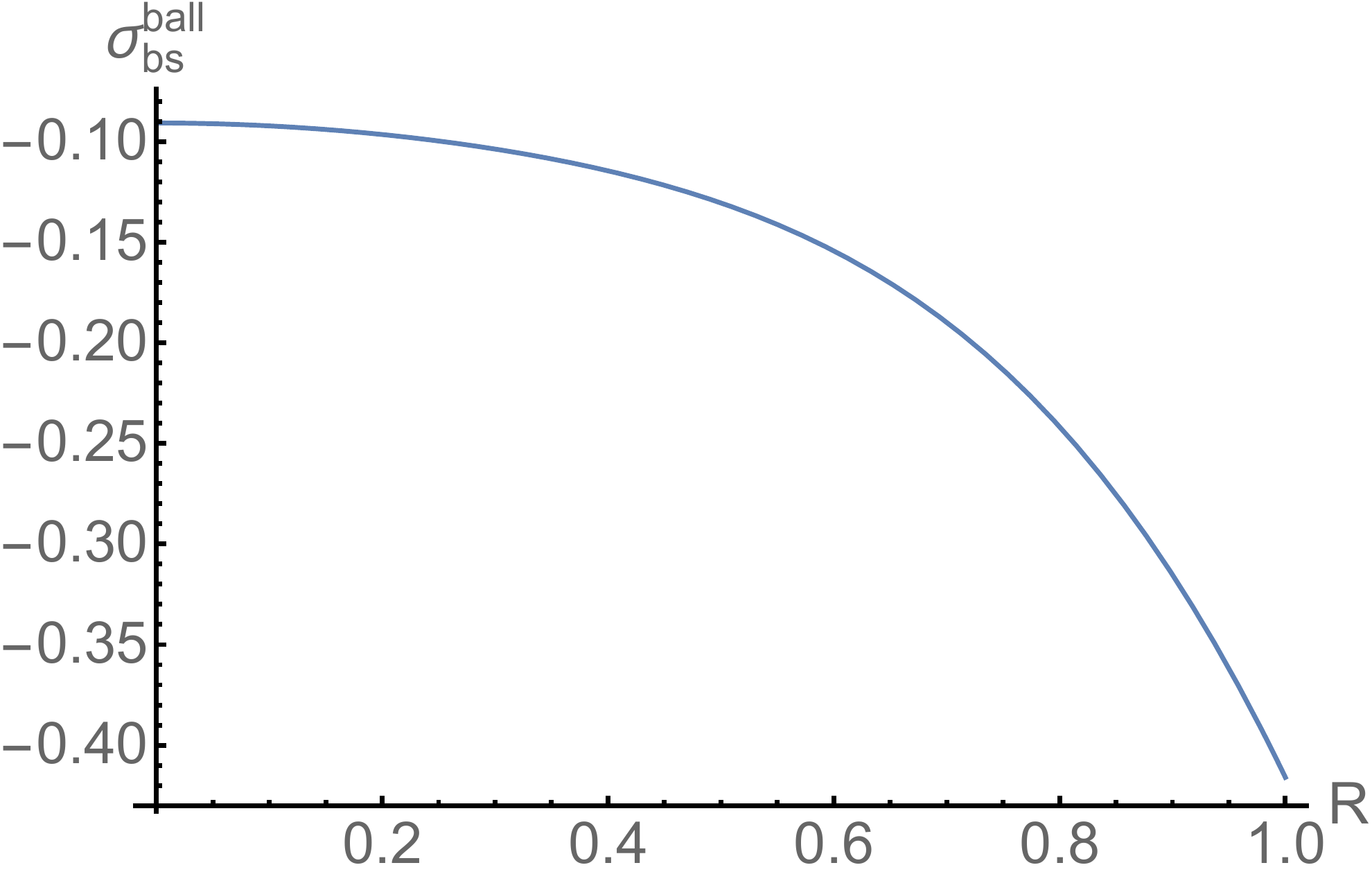}}
\caption{Ballistic part of the backscattering correction in units of $e^2/(2\pi\hbar)$ as a function of $R$ in the absence of dephasing, 
Eq.~(\ref{bs-ball}),
when the logarithmic correction is cut off by the system size $L$, Eq.~(\ref{bs-diff}).}
\label{fig:bs-ball}
\end{figure}

Removing the first term from the integral~\eqref{indefinite_int}, we take the rest in the limits $P_0=0,1$ and obtain the ballistic non-backscattering contribution at $\tau/\tau_\phi\to 0$:
\begin{equation}
\sigma_\text{bs}^\text{ball}=-\frac{e^2}{4\pi^2 \hbar}\frac{1+3R^2}{1+R^2}
\left\{1+R^2-\frac{1}{2}\ln(1+R^2) - \frac{1+2R^2(1-R^2)}{\sqrt{5-4R^2}}
\ln\frac{3+\sqrt{5-4R^2}}{2\sqrt{1+R^2}}\right\} .
\label{bs-ball}
\end{equation}
We emphasize that this ballistic term is a correction to the logarithmic term that was cut off by the
system size, rather than by dephasing.
This result is illustrated in Fig.~\ref{fig:bs-ball}.
In particular, we have
\begin{equation}
\sigma_\text{bs}^\text{ball}=-\frac{e^2}{4\pi^2 \hbar}\left\{
                                                        \begin{array}{ll}
                                                          1-\frac{1}{\sqrt{5}}\ln\frac{3+\sqrt{5}}{2}, & R=0; \\
                                                          4-2\ln 2, & R=1.
                                                        \end{array}
                                                      \right.
                                                      \label{bs-ball-R0R1}
\end{equation}
which gives the values $-0.0144 e^2/\hbar$ at $R=0$ and $-0.066 e^2/\hbar$ at $R=1$.

\subsection*{3. Non-backscattering contribution}

Consider now the diagram for the non-backscattering correction presented in Fig.~\ref{fig:nb}.
This diagram corresponds to the following expression (we neglect $\bm q$ in the current vertices and the disorder correlator):
\begin{multline}
	\sigma_\text{non-bs} = \frac{e^2}{2\pi} \sum_{s,s'}\int\frac{d^2 q}{(2\pi)^2}\int\frac{d^2 k}{(2\pi)^2}\int\frac{d^2 k'}{(2\pi)^2}
\frac{\bm v_{s'}(\bm k') \cdot \bm v_{s}(\bm k) }{ 2}
\left\langle V_{s's}(\bm k',-\bm k) V_{s's}(-\bm k',\bm k)\right\rangle
{\cal C}_{ss'}^{(2)}(\bm k,\bm k',\bm q) \\
	\times
	G^R_{s'}(\bm k' + \bm q)G^R_{s}(\bm k+\bm q)G^A_{s'}(\bm k' + \bm q)G^A_{s}(\bm k+\bm q)
G^R_{s}(-\bm k)G^R_{s'}(-\bm k') + \text{c.c.}
\end{multline}
The superscript $(2)$ of the Cooperon means that this Copperon starts with two impurity lines.

\begin{figure}[h]
\centerline{\includegraphics[width=8cm]{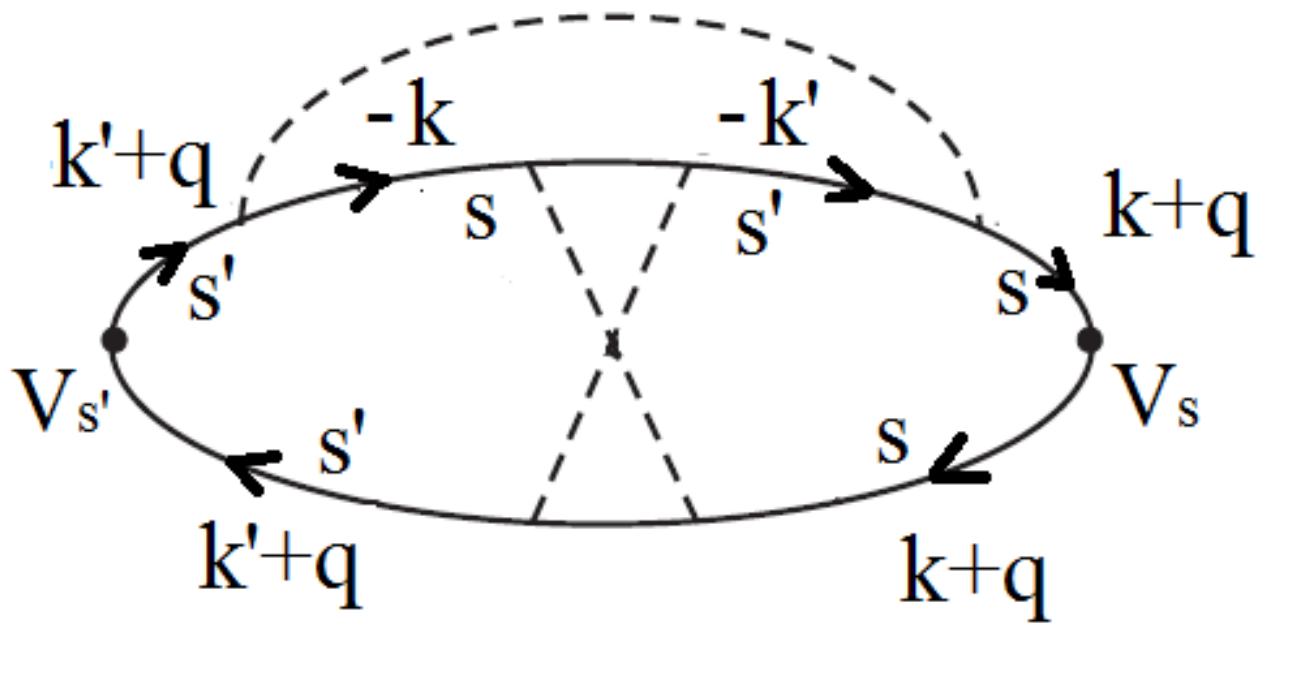}}
\caption{Diagram for the non-backscattering correction (the lowest-order term with a two-impurity Cooperon)}
\label{fig:nb}
\end{figure}

Using the identity (\ref{identity})
for the Green functions at the current vertices and Eq.~(\ref{vs-tr}), we get
\begin{eqnarray}
	\sigma_\text{non-bs} &=& -\frac{e^2 l^2 }{ 4\pi \hbar} \sum_{s,s'}
\int\frac{d^2 q}{(2\pi)^2}\int\frac{d^2 k}{(2\pi)^2}\int\frac{d^2 k'}{(2\pi)^2}
\frac{\tau_{\rm tr}^{(s')} }{ \tau}
 \frac{\tau_{\rm tr}^{(s)} }{ \tau}\frac{\bm k \cdot \bm k' }{ k k'}
\nonumber
\\
&\times&
W_{s's}(\varphi_{\bm k'}-\varphi_{\bm k}-\pi)
{\cal C}_{ss'}^{(2)}(\bm k,\bm k',\bm q)
	G^A_{s}(\bm k+\bm q)
G^R_{s}(-\bm k)G^R_{s'}(-\bm k')G^A_{s'}(\bm k' + \bm q)+ \text{c.c.}
\label{nbGRGA}
\end{eqnarray}
The overall sign has been changed due to ${\rm i}^2$ from the square of the Green function identity.

In Eq.~(\ref{nbGRGA}), we have the following products of retarded and advanced Green's functions:
$G^R_{s'}(-\bm k') G^A_{s'}(\bm k' + \bm q)$ and $G^R_{s}(-\bm k)G^A_{s}(\bm k+\bm q)$.
Since $W_{s's}(\varphi_{\bm k'}-\varphi_{\bm k}-\pi)=W_{s's}(\varphi'-\varphi-\pi)$
and ${\cal C}_{ss'}^{(2)}(\bm k,\bm k',\bm q)={\cal C}_{ss'}^{(2)}(\varphi,\varphi',\bm q)$ do not depend on the absolute values
of $\bm k'$ and $\bm k$, we integrate the products of Green's functions over $|\bm k'|$ and $|\bm k|$ for the fixed values
 of $\varphi=\varphi_{\bm k}-\varphi_{\bm q}$ and $\varphi'=\varphi_{\bm k'}-\varphi_{\bm q}$, similarly to the backscattering case.
Using Eq.~\eqref{PGRGA},
we get
\begin{equation}
	\sigma_\text{non-bs} = -\frac{e^2 l^2 }{ 2\pi V_0^4} \sum_{s,s'} \int\frac{d^2 q}{(2\pi)^2}
\frac{\tau_{\rm tr}^{(s')} }{ \tau_{s'}} \frac{\tau_{\rm tr}^{(s)} }{ \tau_{s}}
\left\langle \cos(\varphi-\varphi') W_{s's}(\varphi'-\varphi-\pi) {\cal C}_{ss'}^{(2)}(\varphi,\varphi',\bm q)
	P(\varphi)P(\varphi')\right\rangle_{\varphi,\varphi'}.
\label{nb-blue}
\end{equation}
The complex conjugated part has simply led to the factor of 2.

The ratio ${\tau_{\rm tr}^{(s)} / \tau_s}$ is given by
\begin{equation}
	\frac{\tau_{\rm tr}^{(s)} }{ \tau_s}=(1-sR)^2 .
\end{equation}
Thus, we arrive at
\begin{eqnarray}
	\sigma_\text{non-bs} &=& -\frac{e^2 l^2 }{ 2\pi \hbar V_0^4} \sum_{s,s'}
\int\frac{d^2q}{(2\pi)^2}(1-sR)^2(1-s'R)^2
	\left\langle W_{ss'}(\varphi-\varphi'-\pi) \cos(\varphi-\varphi')
{\cal C}_{s's}^{(2)}(\varphi',\varphi,\bm q)
	P(\varphi)P(\varphi')\right\rangle_{\varphi,\varphi'}.
	\nonumber
	\\
\label{non-bs-final}
\end{eqnarray}
Note that $W_+(\varphi-\varphi'-\pi)=W_-(\varphi-\varphi')$ and $W_-(\varphi-\varphi'-\pi)=W_+(\varphi-\varphi')$.

It is again instructive to express this correction in terms of the return probabilities as in Ref.~\onlinecite{GKO}. Using Eq.~(\ref{GKOwss'}) and restoring the ratios $\tau_{\rm tr}^{(s)}/\tau$ and $g_s/g_0$,  we write
\begin{equation}
	\sigma_\text{non-bs} = -\frac{e^2}{\hbar} l^2
\sum_{s,s'}
\frac{\tau_{\rm tr}^{(s')} }{ \tau} \frac{\tau_{\rm tr}^{(s)}}{ \tau} \frac{g_s}{g_0} \frac{g_{s'}}{g_0}
	\left\langle \frac{W_{ss'}(\varphi-\varphi'-\pi)}{V_0^2} w_{s's}(\varphi'-\varphi)\cos(\varphi-\varphi')
\right\rangle_{\theta}.
\label{non-bs-return}
\end{equation}
Denoting $\theta=\varphi'-\varphi$ and including the backscattering term (\ref{bs3-w-final}),
we arrive at
\begin{equation}
	\sigma(0) = -\frac{e^2}{2\hbar} l^2
\sum_{s,s'} \frac{g_s}{g_0} \frac{g_{s'}}{g_0}
	\left\langle
\frac{W_{s's}(\pi-\theta)}{V_0^2} w_{ss'}(\theta)
\left[\left(\frac{\tau_{\rm tr}^{(s)} }{ \tau}\right)^2
+\left(\frac{\tau_{\rm tr}^{(s')} }{ \tau}\right)^2+
 2\frac{\tau_{\rm tr}^{(s)} }{ \tau} \frac{\tau_{\rm tr}^{(s')}}{ \tau}\cos\theta\right]
\right\rangle_{\theta},
\label{total-return}
\end{equation}
which generalizes Eq.~(66) of Ref.~\onlinecite{GKO} to the Rashba case.
Indeed, for only one subband, Eq.~(\ref{total-return})
reduces to Eq.~(66) of Ref.~\onlinecite{GKO}:
\begin{equation}
	\sigma(0) = -\frac{e^2}{\hbar} l^2 \left(\frac{\tau_{\rm tr}}{ \tau}\right)^2
	\left\langle
\frac{W(\pi-\theta)}{V_0^2} w(\theta)
\left(1+\cos\theta\right)
\right\rangle_{\theta}.
\label{total-return-TI}
\end{equation}

Let us now return to Eq.~(\ref{non-bs-final}) 
and perform the angular averaging.
We obtain the non-backscattering correction to the conductivity which can be presented in the following form: 
\begin{equation}
\label{zero_field_non-bs_App}
	\sigma_\text{non-bs} = 
	\frac{e^2}{8\pi^2\hbar} 
	(1+R^2)^2 
	\int\limits_0^{1/(1+\gamma)} \frac{dP_0}{P_0^3} 
	{\rm Tr} \left[ (K\Pi K^T + K^T\Pi K) A({\cal I}-A)^{-1}\right],
\end{equation}
with the matrices $A$ and $K$ being introduced in 
the main text. This expression coincides with Eq.~(33) of the main text.

Using Eq.~(\ref{PnP0}), we express 
$\sigma_\text{non-bs}$
at $\gamma=\tau/\tau_\phi=0$ as
\begin{equation}
\sigma_\text{non-bs} = -\frac{e^2 }{ 8\pi^2 \hbar} \int\limits_0^1 dP_0 \mathcal{F}_\text{non-bs},
\qquad
\left.\mathcal{F}_\text{non-bs}\right|_{\gamma=0}
=\frac{\mathcal{H}(P_0,R)}
{(1-P_0)(1+P_0)^3[1+P_0-P_0^2(1-R^2)]},
\end{equation}
where
\begin{eqnarray}
\mathcal{H}(P_0,R)&=&1-2R^2-3R^4+P_0(4+19R^2-2R^4-R^6)
\nonumber
\\
&+&P_0^2R^2(27-6R^2-R^4)-P_0^3(8+7R^2-6R^4-5R^6)+P_0^4(3-5R^2+5R^4-3R^6).
\end{eqnarray}
At $P_0\to 1$ this yields
\begin{equation}
\mathcal{F}_\text{non-bs}\simeq \frac{4R^2}{(1-P_0)(1+R^2)}.
\end{equation}
As a result, the diffusive contribution to the non-backscattering correction reads (here we set $q_\text{min}=1/L$)
\begin{equation}
\sigma_\text{non-bs}^\text{diff} = -\frac{e^2}{4\pi^2 \hbar}\frac{2R^2}{1+R^2}
\int_0^{1-l^2/2L^2} \frac{dP_0}{1-P_0} = -\frac{e^2}{4\pi^2 \hbar}\frac{2R^2}{1+R^2} \ln \frac{2L^2}{l^2}.
\end{equation}
Combining this with $\sigma_\text{bs}^\text{diff}$, Eq.~(\ref{bs-diff}), we obtain the total diffusive correction
with the prefactor independent of $R$, as it should be:
\begin{equation}
\sigma_\text{diff}= \frac{e^2}{4\pi^2\hbar}\left(\frac{1+3R^2}{1+R^2}-\frac{2R^2}{1+R^2}\right) \ln\frac{2L^2}{l^2}
=\frac{e^2}{4\pi^2\hbar}\ln\frac{2L^2}{l^2}.
\label{total-diff}
\end{equation}

To find the ballistic part of the non-backscattering correction, we again remove the singular term
from $\mathcal{F}_\text{non-bs}$ and integrate the rest.
The calculation analogous to the calculation of the backscattering correction yields:
\begin{eqnarray}
\sigma_\text{non-bs}^\text{ball}&=&-\frac{e^2}{4\pi^2 \hbar}
\left\{-\frac{1+4R^2-R^4+2(4-R^2-R^4)\ln 2}{4}\right.
\nonumber
\\
&+&\left.
\frac{4+7R^2-2R^4-R^6}{4(1+R^2)}\ln(1+R^2) + \frac{10+5R^2-7R^6}{2(1+R^2)\sqrt{5-4R^2}}
\ln\frac{3+\sqrt{5-4R^2}}{2\sqrt{1+R^2}}\right\} .
\label{non-bs-ball}
\end{eqnarray}
This result is illustrated in Fig.~\ref{fig:non-bs-ball}.
In particular, we have
\begin{equation}
\sigma_\text{non-bs}^\text{ball}=-\frac{e^2}{4\pi^2 \hbar}\left\{
                                                        \begin{array}{ll}
                                                          -1/4-2\ln2+\frac{\sqrt{5}}{4}\ln\frac{3+\sqrt{5}}{2}, & R=0; \\
                                                          -1+\ln 2, & R=1.
                                                        \end{array}
                                                      \right.
\end{equation}
which gives the values $-0.013 e^2/\hbar$ at $R=0$ and $0.0078 e^2/\hbar$ at $R=1$.
\begin{figure}[ht]
\centerline{\includegraphics[width=10cm]{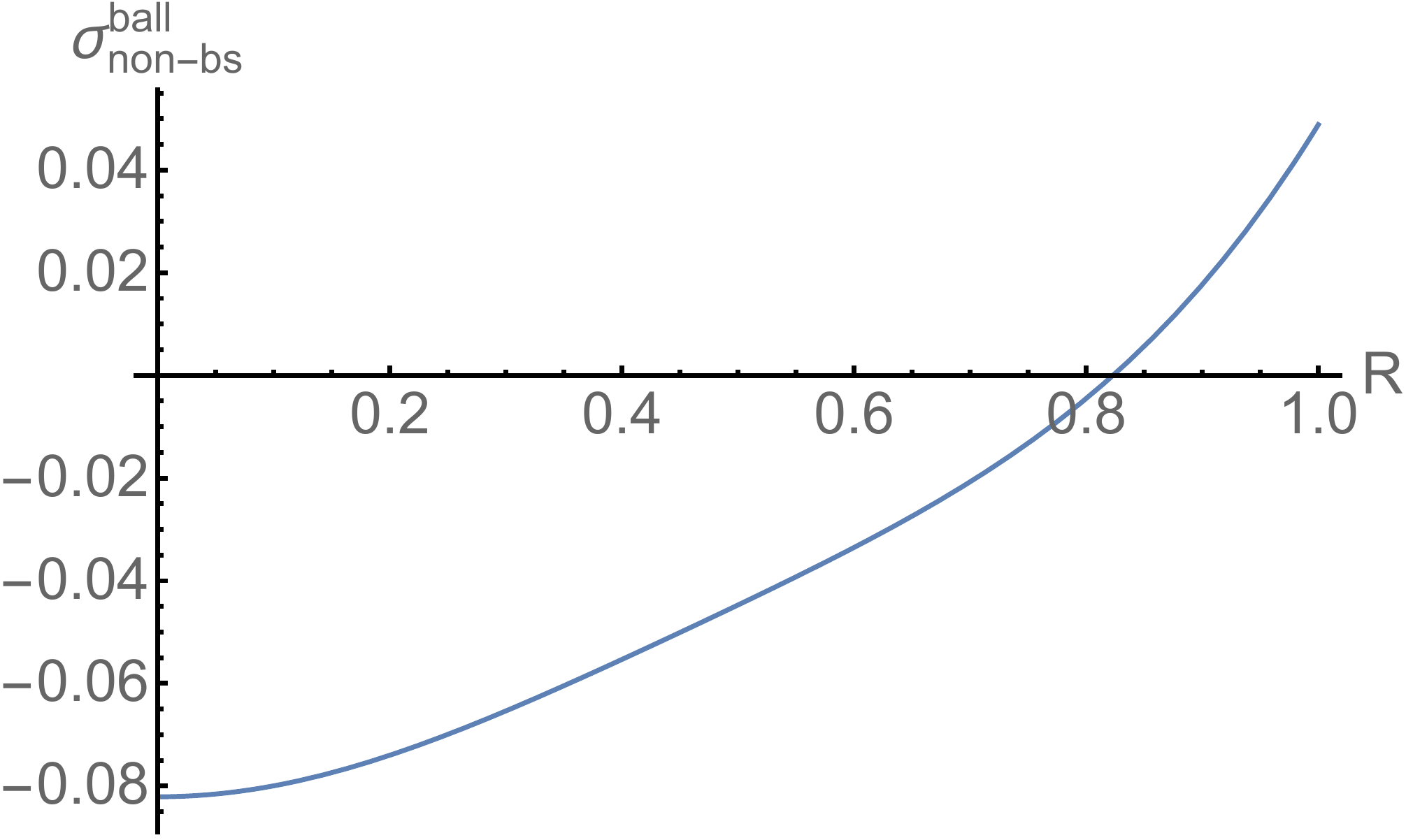}}
\caption{Ballistic part of the non-backscattering correction in units of $e^2/(2\pi\hbar)$ as a function of $R$ in the absence of dephasing, Eq.~(\ref{non-bs-ball}).}
\label{fig:non-bs-ball}
\end{figure}

In order to find the total ballistic contribution at finite $\tau/\tau_\phi$ and arbitrary $R$,
we add up $(1+3R^2)\mathcal{F}_\text{bs}$ and $\mathcal{F}_\text{non-bs}$,
and obtain Eq.~(36) from the main text.

For $R=1$, the integrals over $P_0$ can be easily calculated for arbitrary $\gamma$, since the cubic term $P_0^3$ is multiplied
by $(1-R^2)$ in the denominators of Eqs.~\eqref{angular-bs1},~\eqref{zero_field_non-bs_App}.
Note that above, focusing on the weak dephasing,
we have not included $\tau_\phi$ into the identity (\ref{identity}) for the products $G^RG^A$.
Including $\gamma$ there, we would multiply $\sigma(0)$ by $(1+\gamma)^{-2}$,
and hence at $\gamma\ll 1$ would generate a term $\propto 2\gamma \ln\gamma$ in the conductivity correction.
Restoring this factor we obtain at $R=1$:
\begin{eqnarray}
\sigma_\text{total}(R=1)&=&-\frac{e^2}{4\pi^2(1+\gamma)^2}
\left\{
3+(1+2\gamma)\left[(2+\gamma)\ln\gamma-4(1+\gamma)\ln(2+2\gamma)+(2+3\gamma)\ln(1+2\gamma)\right]
\right.
\nonumber
\\
&+&
\left.
\frac{2(2+10\gamma+13\gamma^2+2\gamma^3)}{\sqrt{4+8\gamma+\gamma^2}}
\ln\frac{2+3\gamma+\sqrt{4+8\gamma+\gamma^2}}{2\sqrt{\gamma+2\gamma^2}}\right\}.
\label{total-R1}
\end{eqnarray}
It is worth stressing once again that the ballistic part of 
Eq.~(\ref{total-R1}) (calculated at finite dephasing) differs from  
the result calculated without dephasing, when the diffusive contribution is cut off by the system size
by the extra term $\ln 2$.
As discussed in the main text,
 this is related to the fact that $2L^2/l^2$ corresponds to $(1+R^2)\tau_\phi/(2\tau)$.

The dependence of $\sigma_\text{total}(R=1)$ on $\tau/\tau_\phi$ at $R=1$ is shown in Fig.~\ref{fig:sigma-tauphi}.
Expanding Eq.~(\ref{total-R1}) for weak dephasing, $\gamma=\tau/\tau_\phi\to 0$, we get:
\begin{equation}
\sigma_\text{total}(R=1, \gamma \to 0)=
\frac{e^2}{4\pi^2}\left[\left(1-\frac{\tau}{\tau_\phi}\right)\ln\frac{\tau_\phi}{\tau}-3+2\ln2+\mathcal{O}(\gamma)\right].
\label{total-R1-g0}
\end{equation}

The total conductivity correction at other representative values of $R$ in Fig.~\ref{fig:sigma-tauphi} has been obtained by numerical integration over $P_0$. For $\tau/\tau_\phi\ll 1$,
this integral can be again easily evaluated analytically for arbitrary $R$ to the leading in $\tau/\tau_\phi$, as described in the main text.
\begin{figure}[ht]
\centerline{
\includegraphics[width=9cm]{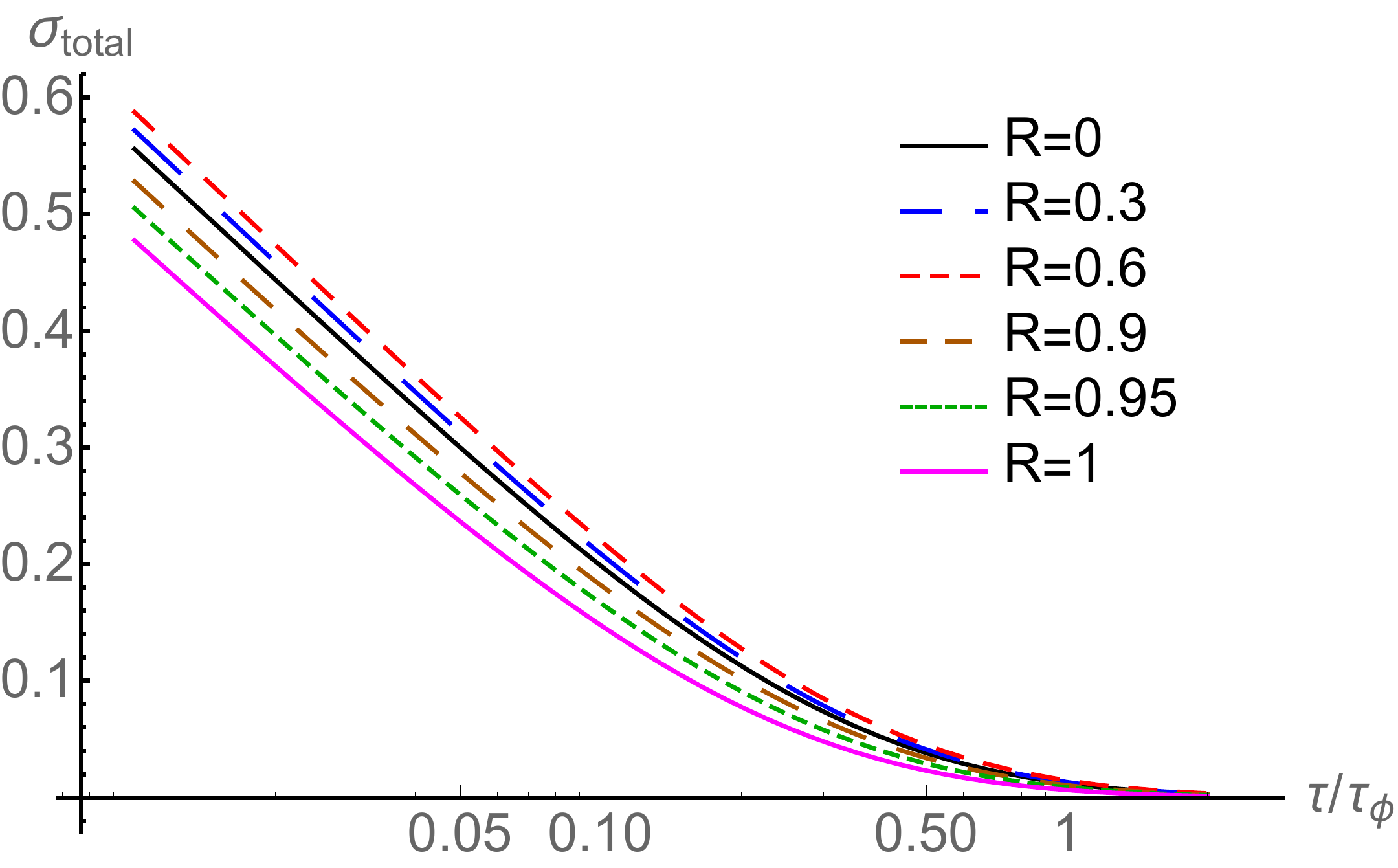}
}
\caption{The total correction in units of $e^2/(2\pi\hbar)$ as a function of $\tau/\tau_\phi$  obtained at $R=1$ from Eq.~(\ref{total-R1}) and obtained by numerical evaluation of $P_0$ integrals from Eq.~(\ref{angular-bs1}) and Eq.~(\ref{zero_field_non-bs_App}) at several representative values of $R$.}
\label{fig:sigma-tauphi}
\end{figure}

\begin{center}
{\bf S2. Dephasing due to Coulomb interaction} 
\end{center}
\label{dephasing}

In this section, we analyze the dephasing processes due to the Coulomb interaction in a system with strong spin-orbit splitting of the spectrum. 
For the calculation of the dephasing rate we will need the expression for the diffusion propagator
that governs the dynamical screening of the interaction. We will also use the diffusons and the polarization
operators for re-deriving the Drude conductivity.

\subsection*{1. Diffuson and polarization operator}

The diffuson equation in the basis of subbands $s=\pm$ is analogous to
Eqs.~(\ref{C_system1}) and (\ref{C_system2}) with the replacement of the disorder correlators
$W_{ss'}(\theta)$ from Eq.~(\ref{corr}) by
\begin{align}
\label{corrD}
\tilde{W}_{+}(\theta)\equiv \tilde{W}_{++}(\theta) = \tilde{W}_{--}(\theta)
=V_0^2\left|\frac{1+{\rm e}^{{\rm i}\theta}}{2}\right|^2
=\frac{V_0^2}{2}(1+\cos\theta)=e^{-{\rm i}\theta}W_+(\theta), \\
\tilde{W}_{-}(\theta)\equiv \tilde{W}_{-+}(\theta) = \tilde{W}_{+-}(\theta)
=V_0^2\left|\frac{1-{\rm e}^{{\rm i}\theta}}{2}\right|^2
=\frac{V_0^2}{2}(1-\cos\theta)=-e^{-{\rm i}\theta}W_-(\theta).
\end{align}
Explicitly, the set of equations for the diffuson has the following form:
\begin{align}
\label{D_system1}
	{\cal D}_{++}(\varphi,\varphi') = \tilde{W}_+(\varphi-\varphi')  + \frac{2\pi\tau}{\hbar}
\left\langle\tilde{W}_+(\varphi-\varphi_1)  g_+  P(\varphi_1) {\cal D}_{++}(\varphi_1,\varphi')
	+ \tilde{W}_-(\varphi-\varphi_1)  g_-  P(\varphi_1) {\cal D}_{-+}(\varphi_1,\varphi')\right\rangle_{\varphi_1}, \\
\label{D_system2}
	{\cal D}_{-+}(\varphi,\varphi') = \tilde{W}_-(\varphi-\varphi')  + \frac{2\pi\tau}{\hbar}\left<\tilde{W}_-(\varphi-\varphi_1)  g_+  P(\varphi_1) {\cal D}_{++}(\varphi_1,\varphi')
	+ \tilde{W}_+(\varphi-\varphi_1)  g_-  P(\varphi_1) {\cal D}_{-+}(\varphi_1,\varphi')\right>_{\varphi_1}.
\end{align}
The solution is expressed in terms of the same matrix $\hat{\cal B}$, Eq.~(\ref{MatrixB}),
that determines the Cooperon:
\begin{equation}
{\cal D}_{ss'}(\varphi,\varphi')=\frac{V_0^2}{2}
\left( \begin{array}{c}
	1\\
s\cos\varphi  \\
s\sin\varphi
	\end{array} \right)^T
	\left(\hat{\cal I}-\hat{\cal B}\right)^{-1}
	\left( \begin{array}{c}
	1\\
s'\cos\varphi'  \\
s'\sin\varphi'
	\end{array} \right),
\label{matrixDss'}
\end{equation}
The structure of the vectors dressing the matrix diffuson is dictated by
the corresponding structure for the disorder correlator (\ref{corrD}):
\begin{equation}
 \tilde{W}_{ss'}(\varphi,\varphi')=\frac{V_0^2}{2}
\left( \begin{array}{c}
	1\\
s\cos\varphi  \\
s\sin\varphi
	\end{array} \right)^T
	\hat{\cal I}
	\left( \begin{array}{c}
	1\\
s'\cos\varphi'  \\
s'\sin\varphi'
	\end{array} \right).
\label{diffWss'}
\end{equation}

We note that the functions $P(\phi)$ instead of dephasing rate include now the frequency.
This is done by the replacement $1/\tau_\phi \to -{\rm i}\omega$ in Eq.~(\ref{Pphi}).
In the diffusion approximation, $ql\ll 1$ and $\omega\tau\ll 1$, we use
\begin{equation}
P_0\simeq 1-\frac{q^2l^2}{2}+{\rm i}\omega\tau, \qquad P_1\simeq {\rm i}\frac{ql}{2},
\qquad
P_0+P_2\simeq 1-\frac{3}{4}q^2l^2+{\rm i}\omega\tau, \qquad
P_0-P_2\simeq 1-\frac{1}{4}q^2l^2+{\rm i}\omega\tau,
\end{equation}
and
introduce the diffusion constant (that has already appeared in the calculation of the diffusive contribution to the conductivity correction)
\begin{equation}
\text{D}=\frac{l^2}{2\tau}(1+R^2).
\label{diff-coeff}
\end{equation}
The diffuson propagators in the diffusive limit of small frequencies and momenta are then written as:
\begin{eqnarray}
{\cal D}_{++}(\varphi,\varphi')\simeq \frac{1}{4\pi g_0\tau}
\left[\frac{1}{\tau}\,\frac{1+{\rm i}qlR(\cos\varphi+\cos\varphi')+(q^2l^2-2{\rm i}\omega\tau)\cos\varphi\cos\varphi'}
{\text{D}q^2-{\rm i}\omega} + 2\sin\varphi\sin\varphi'\right],
\\
{\cal D}_{-+}(\varphi,\varphi')\simeq \frac{1}{4\pi g_0\tau}
\left[\frac{1}{\tau}\,\frac{1-{\rm i}qlR(\cos\varphi-\cos\varphi')-(q^2l^2-2{\rm i}\omega\tau)\cos\varphi\cos\varphi'}
{\text{D}q^2-{\rm i}\omega} - 2\sin\varphi\sin\varphi'\right].
\label{Diff-diff}
\end{eqnarray}
As before, the diffusons ${\cal D}_{--}$ and ${\cal D}_{+-}$ are obtained by $R\to -R$.

Now we can recalculate the Drude conductivity by inserting the diffuson into the RA-bubble with
current vertices. By making use of the identities~\eqref{PGRGA},\eqref{PGAGR}
and directing the external momentum $\bm q$ along $x$-axis,
we write the conductivity as
\begin{eqnarray}
\sigma_D&=&\frac{e^2 v_F^2}{2\pi} \lim\limits_{q\to 0, \omega\to 0}
\sum_{ss'}\int\frac{d\varphi}{(2\pi)}\int\frac{d\varphi'}{(2\pi)}
\cos{\varphi}\cos{\varphi'}
\nonumber
\\
&\times&\left[2\pi g_s\tau P(\varphi)\delta_{ss'}2\pi\delta(\varphi-\varphi')
+(2\pi)^2 g_sg_{s'}\tau^2 P(\varphi)P(\varphi')
{\cal D}_{ss'}(\varphi,\varphi')\right].
\label{Drude1}
\end{eqnarray}
Here the order of limits is usual for the calculation of the DC conductivity:
first the limit $q\to 0$ is taken and only then the frequency is sent to zero.
The first term in Eq.~(\ref{Drude1}) corresponds to the bare bubble, while the second
term with the diffuson produces the vertex corrections (transportization).
Calculating the angular averages over $\varphi$ and $\varphi'$, we see that only the term in the diffusons that contains $\cos\varphi\cos\varphi'$ contributes
 \begin{eqnarray}
\sigma_D=\frac{e^2 v_F^2}{2\pi} \lim_{q\to 0, \omega\to 0}
\sum_{ss'} 2\pi g_s\tau
\left[ \frac{P_0+P_2}{2}\delta_{ss'}
+\frac{g_{s'}}{g_0} \left(\frac{P_0+P_2}{2}\right)^2 \frac{ss'}{2\tau}
\frac{q^2l^2-2{\rm i}\omega\tau}{\text{D}q^2-{\rm i}\omega}\right].
\label{Drude2}
\end{eqnarray}
Taking the limits, we arrive at
\begin{eqnarray}
\sigma_D =
\frac{e^2 v_F^2}{2\pi}
\sum_{ss'} 2\pi g_s\tau
\left[ \frac{1}{2}\delta_{ss'}+(1-s' R)  \frac{ss'}{4} \right]
= e^2(g_++g_-)\text{D} = e^2(g_++g_-)\frac{v_F^2\tau}{2}(1+R^2).
\label{Drude3}
\end{eqnarray}
This result reproduces the Einstein relation for the parallel connection of two conductors
with the same diffusion coefficient (\ref{diff-coeff}) and different densities of states.
At fixed value of $E_F$, the Drude conductivity depends on the Rashba splitting, $\sigma_D\propto 1+R^2$, through the dependence of $\text{D}$ 
on $R$. When the total particle concentration is fixed instead of $E_F$,
the Drude conductivity becomes independent of $R$, see discussion in the main text. 

Let us now calculate the polarization operator.
In the present case of two subbands it can be written as a matrix in $ss'$-space:
\begin{eqnarray}
\Pi_{ss'}(\omega,{\bm q})&=&\langle\tilde{\Pi}_{ss'}(\varphi,\varphi';\omega,{\bm q})\rangle_{\varphi,\varphi'}
\label{Pi-averaged}
\\
\tilde{\Pi}_{ss'}(\varphi,\varphi';\omega,{\bm q})
&=&2\pi g_s\delta_{ss'}[1+{\rm i}\omega\tau P(\varphi)]\delta(\varphi-\varphi')
+{\rm i}\omega \, 2\pi\tau^2 g_sg_s'P(\varphi)P(\varphi'){\cal D}_{ss'}(\varphi,\varphi';\omega,{\bm q}).
\label{Polarization}
\end{eqnarray}
Here the first (diagonal) term contains the contributions of RR and RA bubbles, while
the second term involves the diffuson 
The RA contributions are proportional to $\omega$ in view of the phase-space restriction
for the energy integration that yields the RA combination of Green's function.
In the static limit, the angular average of Eq.~(\ref{Polarization}) reduces to the density of states,
$\Pi_{ss'}=g_s\delta_{ss'}$.
In the limit $q\to 0$, when only the zeroth harmonics $P_0\to (1-{\rm i}\omega\tau)^{-1}$
 survives, the diffuson takes the form
 \begin{eqnarray}
 {\cal D}_{ss'}(\varphi,\varphi';\omega,0)
=\frac{1}{4\pi g_0 \tau}
 \left(1+\frac{1}{-{\rm i}\omega\tau}
 +ss'\cos\varphi\cos\varphi'
 +\frac{2(1-{\rm i}\omega\tau)}{1-2{\rm i}\omega\tau}ss'\sin\varphi\sin\varphi'\right).
 \end{eqnarray}
In the homogeneous limit, the total polarization operator vanishes, as it should
(a homogeneous external potential does not affect the total density):
\begin{eqnarray}
\sum_{ss'}\Pi_{ss'}(\omega,0)
=0
.
\label{Polarization-q0}
\end{eqnarray}
Finally, we extract the Drude conductivity from the polarization operator by means of the
identity
\begin{equation}
\sigma=e^2\lim_{q\to 0,\omega\to 0}\frac{-{\rm i}\omega}{q^2}\Pi(\omega,{\bm q}),
\label{sigmaPi}
\end{equation}
which follows from the continuity equation (again, the limit $q\to 0$ is taken first).
Using the diffusive approximation for the diffuson, Eq.~(\ref{Diff-diff}),
we substitute it into Eq.~(\ref{Polarization}) and calculate the full polarization operator $\Pi=\sum_{ss'}\Pi_{ss'}$ in the diffusive limit:
\begin{equation}
\Pi=g_++g_-+\frac{{\rm i}\omega}{\text{D}q^2-{\rm i}\omega}\frac{1}{2g_0}\left(g_+^2+g_-^2+2g_+g_-\right)
=2g_0\frac{\text{D}q^2}{\text{D}q^2-{\rm i}\omega}.
\label{Polar-diff}
\end{equation}
Substituting this into Eq.~(\ref{sigmaPi}), we recover
$$
\sigma_D=2e^2 g_0 \text{D}=e^2(g_++g_-)\frac{v_F^2\tau}{2}(1+R^2).
$$
%
%
%

\subsection*{2. Calculation of dephasing rate}

We are now in a position to calculate the dephasing rate in Cooperons due to the inelastic electron-electron
scattering~\cite{AAreview}. In fact, dephasing processes in a two-band system are described by the matrix self-energy
$\Sigma^{\phi}_{ss'}$ for Cooperons, see Fig.~\ref{fig:deph}. Moreover, the angular harmonics of the Cooperon
may also be characterized by their own dephasing rates. Further, the inelastic scattering rates calculated using the golden-rule formula for two non-equal subbands in a ballistic system would differ from each other due to the difference in density of states, similarly to the out-scattering rates $1/\tau_s$ for elastic scattering. However, as we will see below, for sufficiently long
trajectories, the main contribution to the dephasing rates becomes isotropic in subband space because of multiple transitions between the bands due to elastic scattering. In what follows, we will focus on the case of sufficiently low temperatures, when $\tau_\phi\gg\tau$. Even in this case, however, for sufficiently strong magnetic field,
$B\sim B_\text{tr}$, the difference between elements of the matrix $\Sigma^{\phi}_{ss'}$ is expected to become substantial. Indeed, since the length of relevant trajectories in the presence of magnetic field is controlled by the magnetic length, in strong fields the ballistic motion along short trajectories with small number of interband scattering dominates the conductivity correction, so that the ``isotropization'' of dephasing does not take place.
At the same time, the main contribution to the suppression of interference in this regime is due to the magnetic field and hence the ``anisotropy'' of the dephasing rate is immaterial for strong $B$. Therefore, when $\tau_\phi\gg\tau$, we can still use a single dephasing rate in the whole range of magnetic fields.

\begin{figure}[ht]
\centerline{\includegraphics[width=12cm]{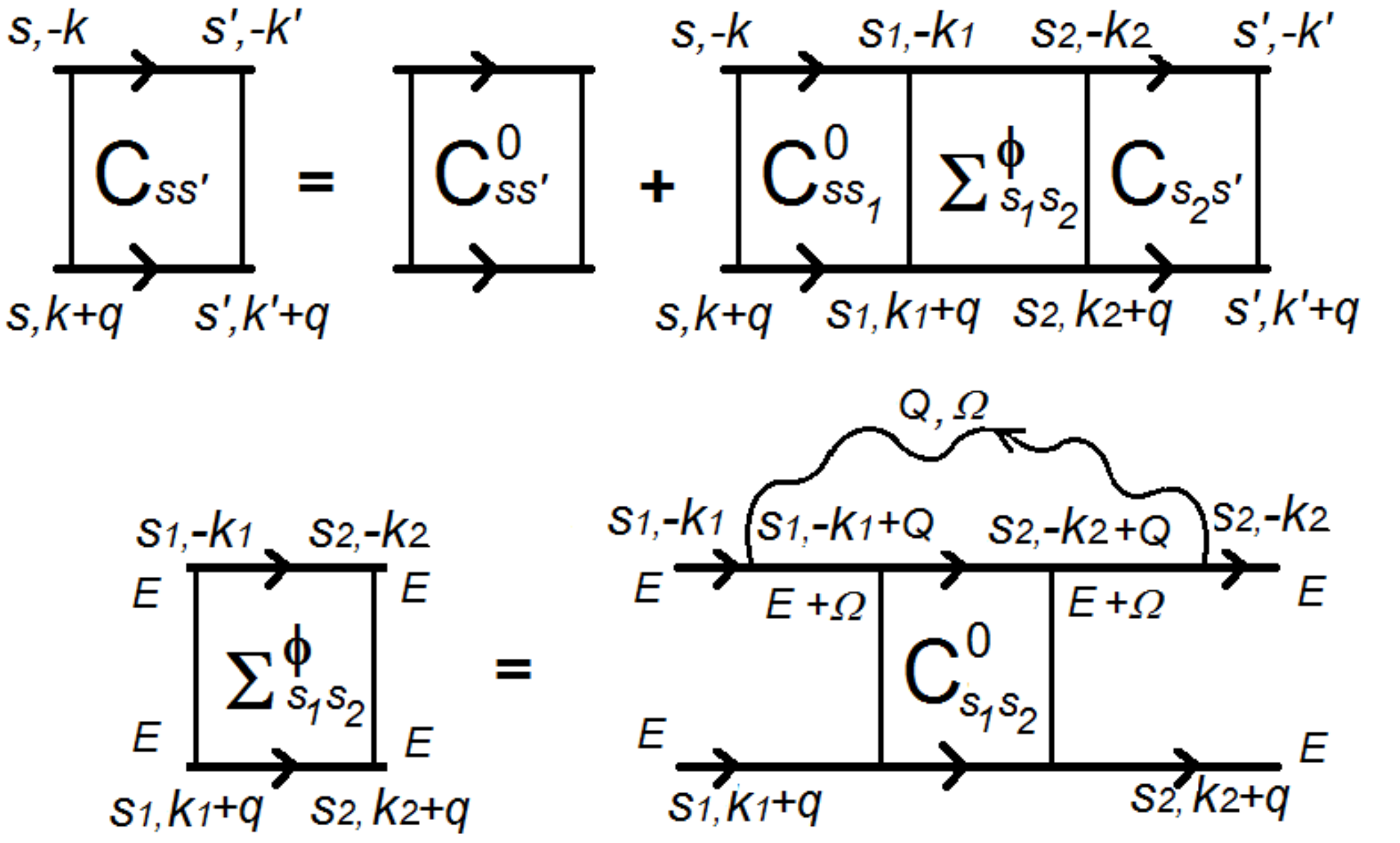}}
\caption{Diagrams describing the dephasing of Cooperons: Bethe-Salpeter equation and the self-energy $\Sigma^\phi_{ss'}$. This self-energy includes all six Green's function as shown in the lower panel.
The full Cooperon and the Cooperon without dephasing ($\gamma=0$) are denoted by
$\mathbf{C}_{ss'}$ and $\mathbf{C}_{ss'}^0$, respectively.
The wavy line denotes the propagator of screened Coulomb interaction.
The self-energy also contains the diagram with the interaction line attached to the
advanced Green functions. }
\label{fig:deph}
\end{figure}

The diagrams describing the effect of inelastic scattering are shown in Fig.~\ref{fig:deph}.
Here we neglect the contribution of diagrams with the interaction line changing the subband
index since the dephasing at low temperatures $\tau_\phi\gg \tau$ is dominated by small
transferred momenta, $Q\ll 1/l \ll |k_+-k_-|$. Furthermore, such terms would involve the non-diagonal
in subbands products of retarded and advanced Green functions at close momenta that are suppressed
in the regime of strong Rashba splitting, $|k_+-k_-|l\gg 1$.

Since the transferred momenta $Q$ are much smaller than the Fermi momenta in the subbands,
we neglect the dressing of interaction vertices by spinor factors.
Then the interaction matrix elements become independent of $ss'$.
Since Cooperons do not depend on the absolute values of momenta $\bm k$ and $\bm k'$,
the self-energies also depend only on the angles of these momenta.

Further, we neglect the dephasing in
the Cooperon in the self-energy and will restore it in the end of the calculation
through the self-consistent cut-off of the integral over the transferred frequency $\Omega$.
We will also disregard the vertex interaction lines connecting the retarded and advanced Green functions in the self-energy of the Cooperon. This approximation is sufficient for the self-consistent
calculation of the dephasing rates since the role of the vertex interaction lines is to regularize the infrared
divergence of the self energy at $Q,\Omega\to 0$. Within the self-consistent calculation, this is done by dephasing itself. Finally, since the characteristic frequencies $\Omega$ are smaller
than $T$, we will use the quasiclassical occupation number $\coth\frac{\Omega}{2T}\to 2T/\Omega$ for the
fluctuations of the electric field created by electron bath and restrict the frequency integration
by $|\Omega|<T$.

The equation for the full Cooperon presented in Fig.~\ref{fig:deph} reads:
\begin{equation}
\mathbf{C}_{ss'}(\varphi,\varphi';{\bm q})
=\mathbf{C}_{ss'}^0(\varphi,\varphi';{\bm q})
+\sum_{s_1s_2}\int\frac{d\varphi_1}{2\pi}\int\frac{d\varphi_2}{2\pi}
\mathbf{C}_{ss_1}^0(\varphi,\varphi_1;{\bm q})
\Sigma^\phi_{s_1s_2}(\varphi_1,\varphi_2,{\bm q})
\mathbf{C}_{s_2s'}(\varphi_2,\varphi';{\bm q}).
\label{Bethe-Salpeter}
\end{equation}
The self-energy is given by
\begin{eqnarray}
&&\Sigma^\phi_{s_1s_2}(\varphi_1,\varphi_2,{\bm q})
=
\int\frac{k_1dk_1}{2\pi}\int\frac{k_2dk_2}{2\pi}
\int_{-T}^T\frac{d\Omega}{(2\pi)}\int\frac{d^2 Q}{(2\pi)^2}
\frac{2T}{\Omega}\text{Im}U(\Omega,{\bm Q})
\mathbf{C}_{s_1s_2}^0(\varphi_1,\varphi_2;\Omega,{\bm Q}+{\bm q})
\nonumber
\\
&&\times
\left[\, G^R_{s_1}( -{\bm k}_1,E)G^R_{s_1}(-{\bm k}_1+{\bm Q},E+\Omega)
G^A_{s_1}({\bm k_1}+{\bm q},E)\right.
G^R_{s_2}( -{\bm k}_2+{\bm Q},E+\Omega)G^R_{s_2}( -{\bm k}_2,E)
G^A_{s_2}({\bm k}_2+{\bm q},E)
\nonumber
\\
&&+
G^R_{s_1}( -{\bm k_1},E)
G^A_{s_1}({\bm k_1}+{\bm q},E)G^A_{s_1}({\bm k_1}+{\bm q}+{\bm Q},E-\Omega)
\left.
G^R_{s_2}( -{\bm k_2},E)
G^A_{s_2}({\bm k_2}+{\bm q}+{\bm Q},E-\Omega)G^A_{s_2}({\bm k_2}+{\bm q},E)\,
\right]
.
\nonumber
\\
\label{self-energy-phi}
\end{eqnarray}
Here the screened interaction involves the total polarization operator:
\begin{equation}
U(\Omega,{\bm Q})=\frac{U_0(\bm Q)}{1+U_0(\bm Q)\sum_{ss'}\Pi_{ss'}},
\label{UOmegaQ}
\end{equation}
where
\begin{equation}
U_0(\bm Q)=\frac{2\pi e^2}{Q}
\end{equation}
is the Fourier transform of the static Coulomb interaction potential
and $\Pi_{ss'}$ is given by Eq.~(\ref{Pi-averaged}).
In the diffusion approximation, $Ql\ll 1,\ \Omega\tau\ll 1$, Eq.~(\ref{UOmegaQ})
reduces to the standard form
\begin{equation}
U(\Omega,{\bm Q})\simeq\frac{1}{\sum_{ss'}\Pi_{ss'}(\Omega,{\bm Q})}
=\frac{1}{2g_0}\frac{\text{D}Q^2-{\rm i}\Omega}{\text{D}Q^2}.
\label{UOmegaQ-diff}
\end{equation}

In principle, the self-energy (\ref{self-energy-phi}) depends on the Cooperon momentum
$\bm q$. After the expansion in small $\bm q$, this dependence renormalizes the mean-free path
(or diffusion coefficient) in the expression for the Cooperon. This corrections are, however, small
in $\tau/\tau_\phi$. Therefore, in order to calculate the dephasing rates for weak dephasing, we set $\bm q=0$ in the self-energy.

Next, we evaluate the integrals over $k_1$ and $k_2$ in Eq.~(\ref{self-energy-phi}).
Using the identity (\ref{identity}) we reduce the products of three Green functions belonging to the same subband
to the products $G^RG^A$ that after the momentum integration yield the functions $P(\varphi)$:
\begin{eqnarray}
\int\frac{k_1dk_1}{2\pi}
G^A_{s_1}({\bm k},E)
G^R_{s_1}(-{\bm k}_1,E)
G^R_{s_1}(-{\bm k}_1+{\bm Q},E+\Omega)
&\simeq&
-{\rm i}\tau \int\frac{k_1dk_1}{2\pi}
 G^R_{s_1}(-{\bm k}_1+{\bm Q},E+\Omega)G^A_{s_1}({\bm k_1},E)
 \nonumber
 \\
&=&-2\pi {\rm i}\tau^2 g_{s_1} P(\varphi;\Omega,\bm Q),
\\
 \int\frac{k_1dk_1}{2\pi}
G^R_{s_1}( -{\bm k_1},E)
G^A_{s_1}({\bm k_1},E)G^A_{s_1}({\bm k_1}+{\bm Q},E-\Omega)
&\simeq&{\rm i}\tau \int\frac{k_1dk_1}{2\pi} G^R_{s_1}(-{\bm k}_1)G^A_{s_1}({\bm k_1}+{\bm Q},E-\Omega)
\nonumber
\\
&=&2\pi {\rm i}\tau^2 g_{s_1} P(\varphi_1;\Omega,\bm Q).
\end{eqnarray}
This leads to
\begin{eqnarray}
\Sigma^\phi_{s_1s_2}(\varphi_1,\varphi_2)
&=&
16 \pi^2 \tau^4 g_{s_1}g_{s_2}
\int_{-T}^T\frac{d\Omega}{(2\pi)}\int\frac{d^2 Q}{(2\pi)^2}
\frac{T}{\Omega}\text{Im}U(\Omega,{\bm Q})
\mathbf{C}_{s_1s_2}^0(\varphi_1,\varphi_2;\Omega,{\bm Q})
P(\varphi_1;\Omega,\bm Q)P(\varphi_2;\Omega,\bm Q)
.
\nonumber
\\
\label{self-energy-phi-1}
\end{eqnarray}
Assuming sufficiently low temperatures, $T\tau\ll 1$,
we notice that in view of $|\Omega|\ll 1/\tau$ the integral over $Q$ will be determined by diffusive
momenta, $Q\ll 1/l$. Therefore, we use Eq.~(\ref{UOmegaQ-diff}) and
the Cooperon
in the diffusive approximation, similar to Eqs.~(\ref{Diff-diff}), keeping only
the leading term in the numerator:
\begin{eqnarray}
\mathbf{C}_{s_1s_2}^0(\varphi_1,\varphi_2)
&\simeq&\frac{e^{{\rm i}(\varphi_2-\varphi_1)}}{4\pi g_0\tau^2}
\frac{s_1s_2}{\text{D}Q^2-{\rm i}\Omega}.
\label{Coop-diff}
\end{eqnarray}
Making use of the diffusion approximation,
we replace the functions $P(\varphi,\Omega,\bm Q)$ by unities.
Then Eq.~(\ref{self-energy-phi-1}) reduces to
\begin{eqnarray}
\Sigma^\phi_{s_1s_2}(\varphi_1,\varphi_2)
&=&
- \tau^2 \frac{g_{s_1}g_{s_2}}{2\pi g_0^2}\,
\int_{-T}^Td\Omega\int QdQ\,
\frac{T}{\text{D}Q^2}\, e^{{\rm i}(\varphi_2-\varphi_1)}
\frac{s_1s_2}{\text{D}Q^2-{\rm i}\Omega}
\label{self-energy-phi-2}
\end{eqnarray}

We note that the phase factor $\exp[{\rm i}(\varphi_1-\varphi_2)]$ in the self energy is cancelled
by the corresponding factors in the adjacent Cooperons. The only phase factor remaining in the full Cooperon
in Fig.~\ref{fig:deph} is then its overall phase factor. Therefore, we can consider the self-energy $\tilde{\Sigma}^\phi_{s_1s_2}$ for the Cooperons defined in Eq.~(\ref{tildeC}).
This self-energy is to the leading order [with the most singular part (\ref{Coop-diff}) of the Cooperon] independent of angles.
We also use the fact that the frequency appears in propagators
only as $-{\rm i}\Omega$ and take the real part of the $\Omega$-integral, restricting the integration
to positive frequencies:
\begin{eqnarray}
\tilde{\Sigma}^\phi_{s_1s_2}
&\simeq&
- \tau^2 \frac{s_1s_2g_{s_1}g_{s_2}}{2\pi \text{D} g_0^2}\,
\int_{0}^T d\Omega \int d(\text{D}Q^2)\,
\frac{T}{(\text{D}Q^2)^2+\Omega^2}\,
=
-  \tau^2 \frac{s_1s_2g_{s_1}g_{s_2}}{g_0^2}\frac{T}{4 \text{D}}\,
\int_{0}^T \frac{d\Omega}{\Omega}.
\label{self-energy-phi-3}
\end{eqnarray}
The integral over $\Omega$ diverges logarithmically in the infrared.
Recalling now that the Cooperon in the self-energy itself contains
the dephasing, we regularize this divergence self-consistently
at $\Omega$ of the order of the dephasing rate. As a result, we get
\begin{equation}
\tilde{\Sigma}^\phi_{s_1s_2}\simeq
-  \frac{s_1s_2g_{s_1}g_{s_2}}{ g_0 }\ \frac{\pi\tau^2}{\tau_\phi},
\label{SE-final}
\end{equation}
where we have introduced
\begin{equation}
\frac{1}{\tau_\phi}=\frac{T}{4\pi g_0 \text{D}} \ln(T\tau_\phi)
\simeq \frac{T}{4\pi g_0 \text{D}} \ln(4\pi g_0 \text{D}),
\label{tau-phi0}
\end{equation}
which is the standard expression for the dephasing rate (note that $4\pi g_0 \text{D}$ is the dimensionless
Drude conductance of the system in units of $e^2/h$).

The Bethe-Salpeter equation (\ref{Bethe-Salpeter}) now reduces to the equation
for the Cooperons $\tilde{\mathbf{C}}$ without the phase factors:
\begin{equation}
\tilde{\mathbf{C}}_{ss'}(\varphi,\varphi';{\bm q})
=\tilde{\mathbf{C}}_{ss'}^0(\varphi,\varphi';{\bm q})
+\sum_{s_1s_2}\int\frac{d\varphi_1}{2\pi}
\tilde{\mathbf{C}}_{ss_1}^0(\varphi,\varphi_1;{\bm q})
\tilde{\Sigma}^\phi_{s_1s_2}
\int\frac{d\varphi_2}{2\pi}
\tilde{\mathbf{C}}_{s_2s'}(\varphi_2,\varphi';{\bm q})
\label{Bethe-Salpeter-1}
\end{equation}
with the self-energy (\ref{SE-final}).
Integrating Eq.~(\ref{Bethe-Salpeter-1}) over $\varphi$, multiplying it by $s g_s$, and performing the summation over $s$ we reduce it to the algebraic equation for the averaged full Cooperon
$${\bar{\mathbf{C}}}_{s'}(\varphi',{\bm q})=
\sum_s s g_s\left\langle\tilde{\mathbf{C}}_{ss'}(\varphi,\varphi';{\bm q})\right\rangle_\varphi.$$
Using
\begin{equation}
\left\langle\tilde{\mathbf{C}}_{ss'}^0(\varphi,\varphi')\right\rangle_\varphi
=\frac{s}{4\pi g_0 \tau d_2}\left[s'\left(1-\frac{P_0+P_2}{2}\right)-P_1 R \cos\varphi'\right]
\end{equation}
and $\sum_s s^2 g_s = 2g_0,$ we 
find
\begin{equation}
{\bar{\mathbf{C}}}_{s'}(\varphi')
=
\frac{1}{2\pi \tau d_2}\left[s'\left(1-\frac{P_0+P_2}{2}\right)-P_1 R\cos\varphi'\right]
\frac{1}{1+\frac{1}{d_2} \left(1-\frac{P_0+P_2}{2}\right) \frac{\tau}{\tau_\phi} }.
\end{equation}
Substituting this back to Eq.~(\ref{Bethe-Salpeter-1}), we arrive at
\begin{eqnarray}
\tilde{\mathbf{C}}_{ss'}(\varphi,\varphi';{\bm q})
&=&\tilde{\mathbf{C}}_{ss'}^0(\varphi,\varphi';{\bm q})
-\sum_{s_1}
\frac{s_1}{4\pi g_0 \tau d_2}\left[s\left(1-\frac{P_0+P_2}{2}\right)-P_1 R\cos\varphi\right]
\pi \tau^2 \frac{s_1g_{s_1}}{g_0 \tau_\phi}{\bar{\mathbf{C}}}_{s'}(\varphi')
\nonumber
\\
&=&
\tilde{\mathbf{C}}_{ss'}^0(\varphi,\varphi';{\bm q})
-
\frac{1}{4 \pi g_0 \tau d_2 }
 \frac{\tau}{ \tau_\phi}
\frac{\left[s\left(1-\frac{P_0+P_2}{2}\right)-P_1 R\cos\varphi\right]
\left[s'\left(1-\frac{P_0+P_2}{2}\right)-P_1 R\cos\varphi'\right]}
{ d_2 + \left(1-\frac{P_0+P_2}{2}\right) \frac{\tau}{\tau_\phi} }.
\nonumber
\end{eqnarray}
In the diffusion approximation $ql\ll 1$, this yields
\begin{eqnarray}
\mathbf{C}_{ss'}(\varphi,\varphi';{\bm q})
&\simeq&
\frac{e^{{\rm i}(\varphi'-\varphi)}}{4\pi g_0\tau^2}
\left\{\frac{ss'+{\rm i}qlR(s'\cos\varphi+s\cos\varphi')}{\text{D}q^2 + \frac{1}{\tau_\phi}}+
2\tau\left[\sin\varphi\sin\varphi'
+
\frac{\frac{\text{D}}{1+R^2}q^2 + \frac{1}{\tau_\phi}}{\text{D}q^2 + \frac{1}{\tau_\phi}}
\cos\varphi\cos\varphi'\right]
\right\}.
\nonumber
\\
\label{full-Coop-diff}
\end{eqnarray}
The last term here is non-singular at $\text{D}q^2\sim 1/\tau_\phi \ll 1/\tau$ and does not
lead to the logarithmic divergence of the interference correction.
The WAL logarithm is therefore cut off by $1/\tau_\phi$ given by Eq.~(\ref{tau-phi0}).

In the experimentally relevant case, it is the electron concentration $n$
that is kept fixed. In this case, the diffusion coefficient
$\text{D}$  and the ratio $\tau_\phi/\tau$
do not depend on $R$, see the main text and (\ref{tau-phi0}):
\begin{equation}
\frac{\tau}{\tau_\phi} =\frac{m T}{2 \pi n } \ln\frac{2\pi n \tau}{m}.
\end{equation}
Then the $T$-dependent diffusive contribution to the interference correction to the conductivity is $R$-independent,
\begin{equation}
\sigma^\text{diff}=\frac{e^2}{4\pi^2 \hbar}\ln\frac{\tau_\phi}{\tau}
\simeq \frac{e^2}{4\pi^2 \hbar}\ln\frac{2\pi n}{m T \ln(2\pi n \tau/m)},
\label{sigma-diff-ne}
\end{equation}
as emphasized in the main text.

%
%
%
%
%
\end{widetext}
\end{document}